\documentclass[12pt]{article}
\usepackage[left=2.5cm, right=2.5cm, top=2.5cm]{geometry}
\usepackage[utf8]{inputenc}
\usepackage{amsfonts}
\usepackage{amsmath}
\usepackage[english]{babel}
\usepackage{textcomp}
\usepackage{graphicx}
\usepackage{subcaption}
\usepackage{booktabs,color}
\usepackage{tabularx}
\usepackage{framed}
\usepackage{bbm}
\usepackage{dsfont}
\usepackage{dirtytalk}
\usepackage{algorithm}
\usepackage{algpseudocode}

\usepackage[autostyle]{csquotes}
\usepackage[export]{adjustbox}
\usepackage{mathrsfs}

\DeclareMathOperator*{\argmin}{argmin}
\DeclareMathOperator*{\argmax}{argmax}
\RequirePackage{filecontents}
\usepackage{hyperref}
\usepackage[noabbrev]{cleveref}
\usepackage{subcaption}
\usepackage[T1]{fontenc}
\usepackage{setspace}
\usepackage{natbib}
\setcitestyle{authoryear,open={(},close={)}}
\usepackage{wrapfig}
\graphicspath{{Users/francesca/Dropbox (Princeton)/Research (private)/sbm/covid/arxiv }}

\usepackage{cite}
\makeatletter
\newcommand{\multiline}[1]{%
  \begin{tabularx}{\dimexpr\linewidth-\ALG@thistlm}[t]{@{}X@{}}
    #1
  \end{tabularx}
}
\makeatother
\renewcommand{\hat}{\widehat}

\newcommand{\bfm}[1]{\ensuremath{\mathbf{#1}}}

   \def\bA{\bfm A}

   \def\bI{\bfm I}

     \def\RR{\mathbb{R}}
   \def\bS{\bfm S}  
     
   \def\bU{\bfm U}

\def\by{\bfm y}

 \def\cG{{\cal  G}}

\newcommand{\bfsym}[1]{\ensuremath{\boldsymbol{#1}}}

            \def\bTheta {\bfsym {\Theta}}

\DeclareMathOperator{\diag}{diag}

\DeclareMathOperator{\rank}{rank}

\begin{document}
\title{The Interplay of Demographic Variables and Social Distancing Scores in Deep Prediction of U.S. COVID-19 Cases
\thanks{Tang and Feng contribute equally to this work.}
}
\author{Francesca Tang\thanks{Department of Operations Research and Financial Engineering, Princeton University, Princeton, NJ (email: frtang@princeton.edu).}\text{                } $\,\,\,\,\,\,\,\,$ Yang Feng\thanks{Department of Biostatistics, New York University, New York City, NY (yang.feng@nyu.edu).} \text{                } $\,\,\,\,\,\,\,\,$ Hamza Chiheb\thanks{New York City, NY (hamza.chiheb@gmail.com).}  \text{                }$\,\,\,\,\,\,\,\,$ Jianqing Fan\thanks{Department of Operations Research and Financial Engineering, Princeton University, Princeton, NJ (jqfan@princeton.edu).} \\
}

\maketitle

\doublespacing
\begin{center}
\section*{Abstract}
\end{center}

With the severity of the COVID-19 outbreak, we characterize the nature of the growth trajectories of counties in the United States using a novel combination of spectral clustering and the correlation matrix. As the U.S. and the rest of the world are experiencing a severe second wave of infections, the importance of assigning growth membership to counties and understanding the determinants of the growth are increasingly evident. Subsequently, we select the demographic features that are most statistically significant in distinguishing the communities. Lastly, we effectively predict the future growth of a given county with an LSTM using three social distancing scores. This comprehensive study captures the nature of counties' growth in cases at a very micro-level using growth communities, demographic factors, and social distancing performance to help government agencies utilize known information to make appropriate decisions regarding which potential counties to target resources and funding to. 

\textbf{Keywords:} COVID-19, Stochastic Block Model, Spectral Clustering, Community Detection, Machine Learning,  Neural Network.

\section{INTRODUCTION}\label{intro}

The recent infectious disease (COVID-19) caused by severe acute respiratory syndrome coronavirus 2 (SARS-CoV-2) has overtaken the world as the largest pandemic we have seen in decades. The World Health Organization (WHO) labeled it a pandemic on 03/11/2020, with a total of more than 85 million cases and more than 1.84 million deaths worldwide as of 01/04/2021. 

Forecasting the growth of confirmed cases and the locations of future outbreaks has been a persistent challenge in the public health and statistical fields. With the gravity and urgency of the global health crisis, many recent works including \citet{Kucharski20} and \citet{Peng20} have attempted to model the growth in cases in various countries. Most of the literature on statistical modeling of the data focuses on the reproduction number. However, this value is constantly evolving and is not always a valuable measurement to build prediction models with. \citet{hong2020estimation} proposed a Poisson model with time-dependent transmission and removal rates to estimate a
time-dependent disease reproduction number. \citet{betensky2020accounting} studied the impact of incomplete testing on the estimation of dynamic doubling time.  Ultimately, we need to examine the underlying features contained in the time series data in order to extract valuable insights into the unique nature of the spread of COVID-19. As the number of deaths is at least a two-week lagging indicator compared to the number of confirmed cases, we only look at the latter. More importantly, the matrix of the number of deaths per county would be very sparse at the initial stage, making any analysis more difficult. Our goal is to characterize and project the disease progression given the limitations of public data from a theoretical yet practical perspective. 

Stochastic block models (SBMs), first developed by \citet{Holland83}, has long been studied as a powerful statistical tool in community detection, where the nodes or members are partitioned into latent groups. SBMs have been employed to study social networks \citet{Wasserman87}, brain connectivity \citet{Rajapakse17}, protein signaling networks \citet{CB06}, and many others. Under an SBM, the nodes within the same group usually have a higher probability of being connected versus those from different groups. The difficult task is to recover these connectivities and the communities based on one observation, which in our case, is a snapshot of the changes in the number of cases up to the most recent time point. In more recent years, spectral clustering \citep{BXK11,Rohe11,jin15,Lei15} has arisen as one of the most popular and widely studied approaches to recover these communities. Conventional spectral clustering algorithms mostly involve two steps: eigen-decompose the adjacency or Laplacian matrix of the data and then apply a clustering algorithm, such as k-means, to the eigenvectors that correspond to the largest or smallest eigenvalues. There is extensive literature on such procedures, for instance \citet{von07}, 
\citet{Ng01}, \citet{abbe2017community}, etc.

In this study, we introduce the unique procedure of conducting spectral clustering on the sample Pearson correlation coefficient matrix directly and compare its clusters to the standard Laplacian embedding. 
This complements \citet{BGL18}'s approach based on a latent covariance model on financial return data. 
\citet{Gilbert20} used agglomerative clustering, an unsupervised learning method, on preparedness and vulnerability data in African countries using self-reported reports of capacity and indicators. While a comprehensive study, it only considers the possible exposures to travelers from China. Using a different dataset, \citet{Hu20} clustered the data from China by implementing a simple $k$-means clustering directly on various features of the provinces/cities and not on the eigenvectors of the correlation matrix. It also doesn't take into account possible explanatory features that aren't directly related to the number of cases and fails to predict provinces that have yet to have cases. The data processing of some existing approaches also do not standardize and shift the data in a way that aligns with the nature of COVID-19. 

Once the communities are found, the subsequent part uncovers the statistically significant demographic features, pre-existing in the counties, that could largely explain a county's community membership. Most of the existing research on salient demographic information focus on age-related features and the presence of co-morbidities or underlying health conditions e.g. \citet{Dowd20} and \citet{Lippi20}. In reality, what influences how the disease progresses in a county is most likely a confluence of variables, and not one or two prevailing ones. Some studies also examine how various demographic determinants affect how well a county carries out social distancing \citep{Im20}, but offers little or no connection to the nature of the growth curve. 

\begin{figure}
  \centering
  \includegraphics[width=85mm,height=85mm]{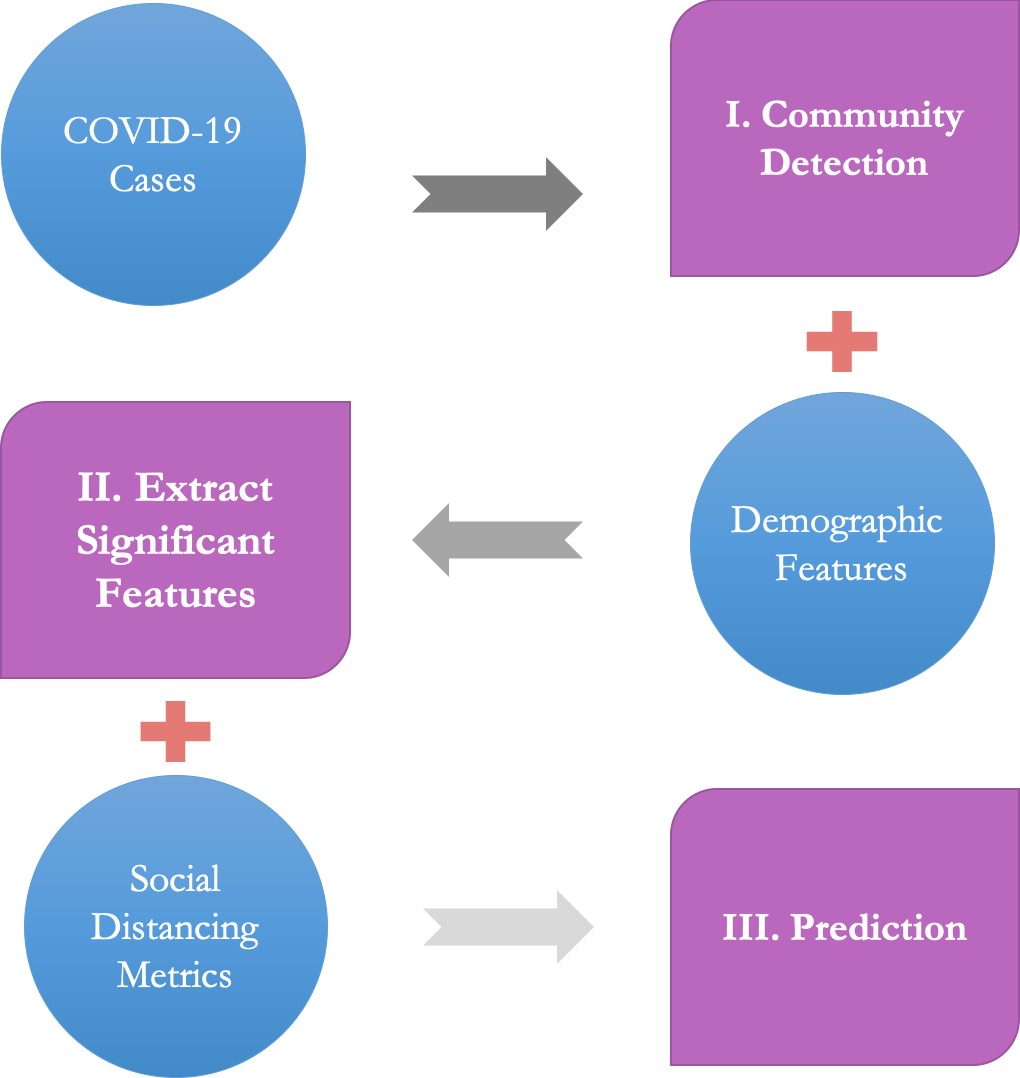}
  \caption{\footnotesize Pipeline of this study's three-part analysis of COVID-19. COVID-19 time series data is first used to perform community detection, clustering counties into several communities. Then, demographic features are incorporated to extract the most significant features that distinguish the growth communities. Finally, social distancing metric time series are added to the results to the previous two parts to carry out the prediction of COVID-19 cases for new counties.}
  \label{fig:flowchart}
  \vspace{-0.5cm}
\end{figure}

The extracted variables from the feature analysis part are then used in conjunction with time series of social distancing scores from \citet{Unacast} to fit a recurrent neural network, ultimately predicting the progression of confirmed cases in a given county. It is important to note that for this prediction section, we use the period from the start of the pandemic until 07/20/2020 as this traces the first large spike in cases in the U.S. and a subsequent plateau. This gives a long enough time series sample and to include much more recent data would include the second large wave of the pandemic, which is counter to the objective of capturing the growth trajectory of a county's peak and fall. Unacast has created a scoreboard of social distancing measures with mobile device tracking data, where a device is assigned to a specific county based on the location the device spent the most amount of time in. The neural network prediction takes these static, inherent county variables, community membership (the clustering results), and social distancing data to predict the future growth of confirmed cases. There have been several studies that predict, estimate, or model the growth curve of the disease, including \citet{Fanelli20} on the cases in Italy, \citet{Roda20} on the cases in Wuhan, and \citet{Liu20} on the cases in China. In addition, deep learning has been applied to COVID-19 research, such as \citet{Wang20} that detect positive cases through chest scans. Other studies such as  \citet{Zheng20} investigate when patients are most infectious by using a deep learning hybrid model and  \citet{Yang20} similarly combines the epidemiological SIR model with an LSTM network. However, the literature on COVID-19 still lacks any comprehensive approach on a county-level that creates a throughline of the pandemic: historical growth curve of confirmed cases, characterization of this growth via clustering, the significant explanatory demographic features, and finally, social distancing measures that give insight into the nature of the future growth trajectory, as displayed in Figure \ref{fig:flowchart}. Table \ref{Tab:partoverview} also contains the specific time period, the number of counties $N$, and the data source used for each part of the paper (Part I: community detection, Part II: extraction of significant features, Part III: prediction) as outlined in Figure \ref{fig:flowchart}.

\begin{table*}[h!]
\centering
\footnotesize
\begin{tabular}{@{}lllclllclll@{}}\toprule 
	& Time Period(s) & No. of counties $N$ & Source\\ \midrule
Part I & 1/22/20 - 4/17/20 and 5/10/20 - 7/10/20&	950 & Johns Hopkins CSSE\\
Part II & 1/22/20 - 4/17/20 and 5/10/20 - 7/10/20&	633& ACS (c/o Data Planet)\\
Part III & 2/25/20 - 7/10/20&	627& Unacast\\
\bottomrule
\end{tabular}
\caption{\footnotesize Time period, number of counties, and data source used for each part of the paper.}
\label{Tab:partoverview}
\end{table*}

\begin{center} \section{COMMUNITY DETECTION} \end{center} \label{CD}
The first part of this paper finds potential communities among the counties, in which  clusters share similar growth patterns. To accomplish this, two fundamental concepts are necessary: the stochastic block model and spectral clustering. The former is a generative model through which  community memberships were formed and the latter is a methodology often utilized to recover these memberships. Compared to traditional clustering methods, spectral clustering has shown to be  effective in both statistical and computational efficiency \citep{abbe2017community,abbe2020entrywise}. Our approach applies spectral clustering to the correlation matrix, instead of the commonly used adjacency matrix or Laplacian matrix. The goal is to recover the county membership matrix embedded in the correlations of each county's logarithmic daily cumulative number of cases. 

\subsection{\textit{Correlation matrix vs. adjacency matrix}} \label{corrvsadj}
For each county, consider a daily time-series of the cumulative number of confirmed cases, where we use curve registration (the time origin is set as the day on which the number of cases exceeds 12 for a particular county).  This curve registration is important as it takes into account the fact that counties may have different COVID-19 outbreak starting times. We denote $w_{i,t} = \log({x_{i,t}})$ as the logarithmic cumulative number of cases of county $i$ on the $t$-th day since the county hit 12 or more cases. Then, we use the Pearson correlation as a similarity measure, defined as

\begin{equation}\label{eq:corr}
    R_{ij} = \frac{\sum_{t=1}^{T_{ij}} (w_{i,t}-\bar{w}_i)(w_{j,t}-\bar{w}_j)}{\sqrt{\sum_{t=1}^{T_{ij}} (w_{i,t}-\bar{w}_i)^2} \sqrt{\sum_{t=1}^{T_{ij}} (w_{j,t}-\bar{w}_j)^2}},
\end{equation}
where $T_{ij}=\min(T_i,T_j),$ with $T_i$ and $T_j$ being the number of days county $i$ and county $j$ has 12 or more cases, respectively. The sample correlation $\boldsymbol{R} \in \RR^{n \times n}$ would then contain the pairwise correlations among all $n$ counties. The logarithmic cumulative case counts are used to align with the exponential growth pattern implied by popular epidemic models. For example, we could distinguish between a faster exponential growth function such as exp$(2t)$ and a slower growth function exp$(t/2)$.

Another commonly used network representation is the adjacency matrix $\boldsymbol A$, which shows whether two counties are connected and is often constructed based on a similarity measure like Pearson correlation or a mutual information score. If the graph is undirected, where each edge that connects two nodes is bidirectional, $\boldsymbol{A}$ is symmetric. The two most common types of similarity graphs are the $\epsilon$-neighborhood graph and the $k-$nearest neighbor graph. As we're using sample correlation as the similarity measure, an $\epsilon$-neighborhood adjacency $\boldsymbol{A}_1$ is defined as follows:
\begin{equation}
    {(A_1)_{ij}}=
    \begin{cases}
      1, & \text{if $ R_{ij} \geq 1-\epsilon$},\ \\
      0, & \text{otherwise.}
    \end{cases}
\end{equation}
 
\noindent A $k-$nearest neighbor adjacency $\boldsymbol{A}_2$ is defined as follows:
\begin{equation}
    (A_2)_{ij}=
    \begin{cases}
      1, & \text{if county $i$ is among $j$'s $k$ nearest neighbors}\\
      &\text{or if county $j$ is among $i$'s $k$ nearest neighbors,}\ \\
      0, & \text{otherwise,}
    \end{cases}
\end{equation}
where the nearest neighbors are found with respect to $R_{ij}$.  

Depending on the parameters $\epsilon$ and $k$ one chooses for $\boldsymbol{A}_1$ and $\boldsymbol{A}_2$, respectively, a significant amount of information could be lost in the process because of the thresholding operation. However, this operation also filters out many spurious correlations.
Unlike the sparse $\boldsymbol{A}_1$ and $\boldsymbol{A}_2$, $\boldsymbol{R}$ retains all of the pairwise similarities between counties, which would shed more light on the within-group and between-group relationships.

\subsection{\textit{Stochastic Block Model (SBM)}} \label{SBM}
The matrices $\boldsymbol{R}$, $\boldsymbol{A}_1$, and $\boldsymbol{A}_2$ are critical because they can help us recover $\boldsymbol{\Theta}$, an $n\times K$ membership matrix that reflects which community each county belongs to, where $K$ is the number of communities. Letting ${Z_i} \in \{1,...,K\}$ be the community that  county $i$ belongs to,  the $i^{th}$ row of $\boldsymbol{\Theta}$ has exactly one 1 in column $Z_i$ (the community that county $i$ belongs to) and 0 elsewhere. We estimate $\boldsymbol{\Theta}$ under an SBM, where the probability two counties are connected only depends on the membership of these two counties. An SBM  denoted by $\mathbb{G}(n,\boldsymbol{B},\boldsymbol{\Theta})$ as $n$ nodes, $K$ communities, and is parameterized by $\boldsymbol{\Theta}$ and $\boldsymbol{B}$, the $K \times K$ symmetric connectivity matrix. Essentially, $\boldsymbol{B}$ contains the inter- and intra-community connection probabilities: the probability of an edge between counties $i$ and $j$ is $B_{Z_i Z_j}$. 

The objective is to obtain an accurate estimation $\boldsymbol{\hat{\Theta}}$ of $\boldsymbol{\Theta}$ from an observed adjacency matrix $\bA$ that is modeled as  $\mathbb{G}(n,\boldsymbol{B},\boldsymbol{\Theta})$.  This yields an recovery of the partitions $G_k := \{i: Z_i = k, i = 1,...,n\}$ by $\hat{G}_k = \{i: \hat Z_i = k, i = 1,...,n\}, k=1,...,K$, with an ambiguity of permutation of clusters, where $\hat Z_i $ indicates the location of 1 in the $i^{th}$ row of $\hat \bTheta$. The population matrix $\boldsymbol{P} \in \mathbb{R}^{n \times n}$, where $P_{ij}$ is the probability that counties $i$ and $j$ are connected, is naturally expressed as $\boldsymbol{P} = \boldsymbol{\Theta B \Theta}^T$. 

\subsection{\textit{Spectral Clustering}} \label{spectral}
Spectral clustering has been a popular choice for community detection \citep{Rohe11,jin15,Lei15}. The central idea is to relate the eigenvectors of the observable adjacency matrix $\boldsymbol{A}$ to those of $\boldsymbol{P} = \boldsymbol{\Theta B \Theta}^T$, which is not observed. This is accomplished by expressing $\boldsymbol{A}$ as a perturbation of its expected value: $\boldsymbol{A}=\mathbb{E}[\boldsymbol{A}]+(\boldsymbol{A}-\mathbb{E}[\boldsymbol{A}]).$ If we treat $\mathbb{E}[\bA]$ as the signal part and $\boldsymbol{A}-\mathbb{E}[\boldsymbol{A}]$ as the noise, we connect the eigenvectors of $\boldsymbol{A}$ and $\boldsymbol{P}$ using $\mathbb{E}[\boldsymbol{A}]=\boldsymbol{P}-\diag(\boldsymbol{P})$. Noting $\rank(\boldsymbol{P})=K$, 
letting $\boldsymbol{U}_{n\times K} = [\boldsymbol{u}_1,...,\boldsymbol{u}_K]$ be the eigenspace spanned by the $K$ nonzero eigenvalues of $\mathbb{E}[\boldsymbol{A}]$, then columns of $\bU$ span the same linear space as those spanned  by the columns of $\boldsymbol{P}$ (ignoring  $\diag(\boldsymbol{P})$).  Additionally, $\boldsymbol{P}$ has the same column space as $\boldsymbol{\Theta}$. Now, letting $\hat{\bU}$ be the eigenspace corresponding to the $K$ largest absolute eigenvalues of $\boldsymbol{A}$, then $\hat{\bU}$ is a consistent estimate of $\bU$ or the column space of $\bTheta$, under some mild conditions.  To resolve the ambiguity created by rotation, the $k$-mean algorithm is applied to the normalized row of $\bU$ to identify membership of communities \citep{Rohe11,Lei15}.

Instead of examining the eigenvalues of $\boldsymbol{A}$, spectral graph theory has long studied graph Laplacian matrices as a tool of spectral clustering. The symmetric Laplacian matrix is defined as follows: letting $\boldsymbol{D} = \diag(d_1,...,d_n)$ be the diagonal degree matrix  where $d_i = \sum_{j=1}^{n} A_{ij}$, then a popular definition of a normalized, symmetric Laplacian matrix is $\boldsymbol{L} =\bI -  \boldsymbol{D}^{-1/2}\boldsymbol{AD}^{-1/2}$. When clustering with $\boldsymbol{L}$, one takes the eigenvectors corresponding to the smallest  eigenvalues in absolute value. 

In our context, $\boldsymbol{A}$ can be taken as either $\boldsymbol{A}_1$ or $\boldsymbol{A}_2$ as outlined in \autoref{corrvsadj}. As there are no exact rules in choosing the parameters $\epsilon$ and $k$ of $\boldsymbol{A}_1$ and $\boldsymbol{A}_2$, respectively, clustering with $\boldsymbol{L}$, which depends on the adjacency matrix, maybe less than ideal. It is also an added, often computationally cumbersome step. Instead, we cluster directly on the similarity matrix $\boldsymbol{R}$, the sample correlation matrix. Algorithm~\ref{alg:Algo1} delineates the detailed steps of this approach. The classic spectral clustering procedure with $\boldsymbol{L}$ used as 
a benchmark is outlined in the Supplementary Material. 

\begin{algorithm}
  \caption{Spectral clustering on correlation matrix}
  \textbf{Input} Sample correlation matrix $\boldsymbol{R} \in \mathbb{R}^{n \times n}$ and the number of clusters $K$.
  \begin{algorithmic}[1]
        \State Compute the largest $K$ eigenvectors in absolute value $\boldsymbol{u}_1, . . . , \boldsymbol{u}_K$ of $\boldsymbol{R}$ and construct $\boldsymbol{\hat{U}} \in \mathbb{R}^{n \times K}$ be the matrix with the eigenvectors as columns.
        \State Normalize rows of $\boldsymbol{\hat{U}}$ to have unit norm to get $\boldsymbol{\hat{U}}_{norm}$.
        \State Cluster the rows of $\boldsymbol{\hat{U}}_{norm}$ with $k$-means.
  \end{algorithmic}
  \textbf{return} Partition $\hat{G}_1, . . . , \hat{G}_K$ of the nodes. \label{alg:Algo1}
\end{algorithm}

\begin{wrapfigure}{R}{12cm}
 \centering
 \vspace*{-0.3in}
  \includegraphics[width=12cm, height =5cm]{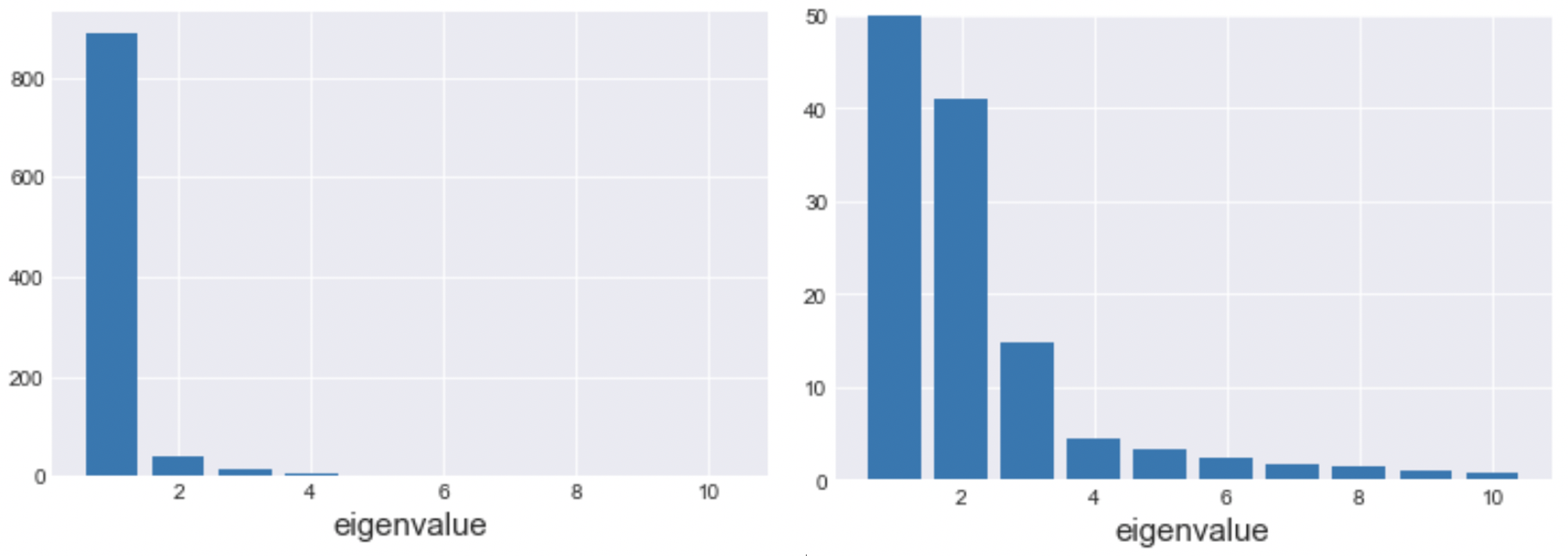}
  \vspace*{-0.3in}
   \caption{The left panel is the 10 largest eigenvalues of $\boldsymbol{R}$. The right panel has the same 10 eigenvalues but zoomed-in on the y-axis.} \label{fig:eig}
   \vspace*{-0.3in}
\end{wrapfigure}

There are several methods for choosing the number of spiked eigenvalues in the context of factor models:  scree-plot, eigen-gap, eigen-ratio, adjusted correlation thresholding. 
As our method involves correlations, we apply the adjusted correlation method in \citet{fan2020estimating}.
This method leads to $K=2$, which roughly divides the counties into faster or slower growth communities. It also agrees with the choice where we maximize the eigen-gap.

\subsection{\textit{Clustering procedure}} \label{clustering procedure}
To compare and visualize the eigenvalues of $\boldsymbol{R}$, Figure \ref{fig:eig}'s left panel displays how dominant the first eigenvalue is compared to the rest.

\begin{wrapfigure}{R}{12cm}
  \centering
  \vspace*{-0.3in}
  \includegraphics[width=12cm,height =6cm]{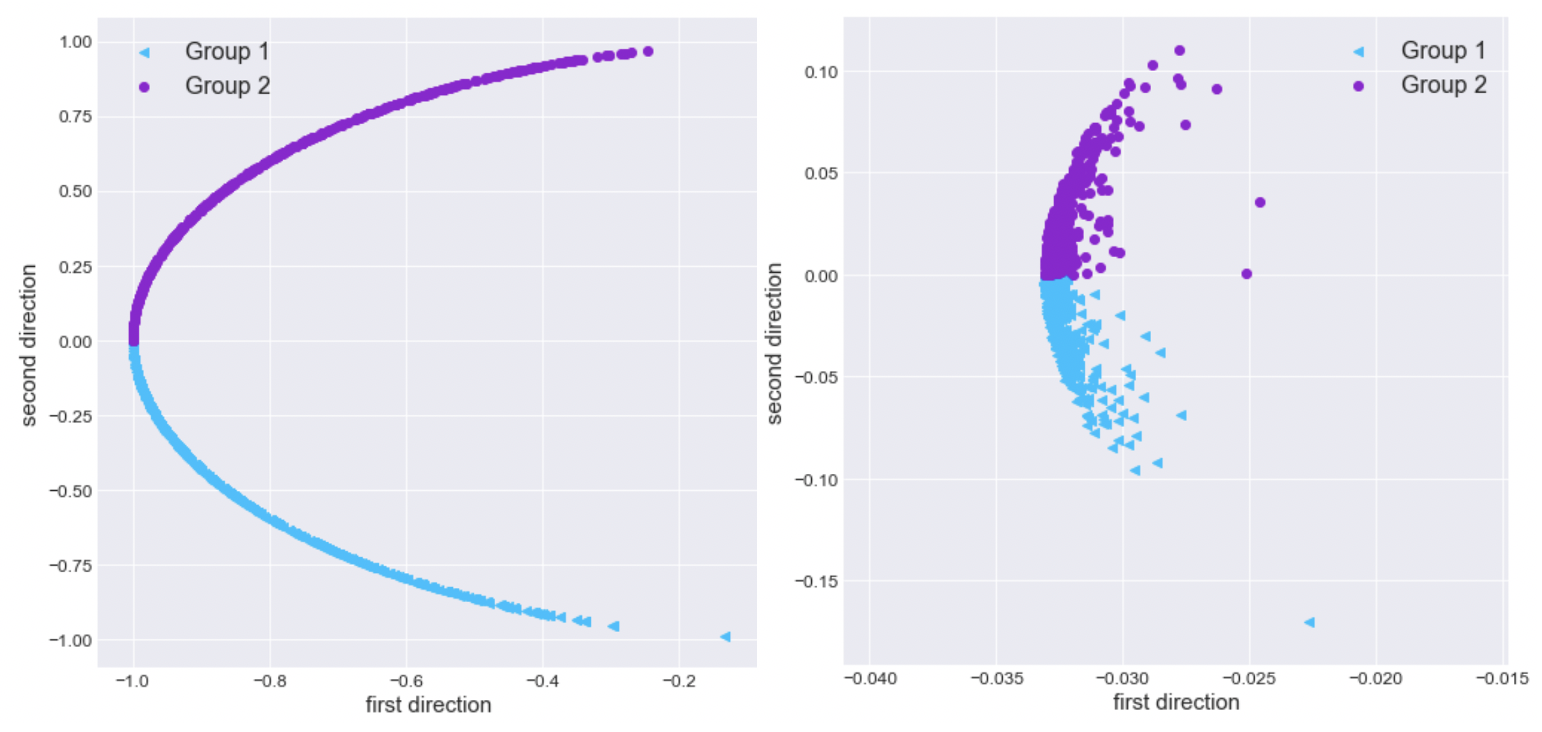}
   \caption{The left panel is the first two unit-norm normalized eigenvectors of $\boldsymbol{R}$ and the corresponding clusters, Group 1 in blue and Group 2 in purple. The right panel depicts the same two clusters but in the 
   two un-normalized eigenvectors.}\label{fig:eigenvectors}
\end{wrapfigure}

Figure \ref{fig:eigenvectors} is a visualization of the first two eigenvectors of $\boldsymbol{R}$ and the linear separation that the algorithm partitioned all the counties into. The left panel is with unit norm normalization and the right is without the normalization. The result of essentially using the signs of the components in the second eigenvector to cluster reminiscences the work by \citet{abbe2020entrywise} with strong theoretical support.
From now on, all clustering analysis will be based on the unit-norm normalization of the eigenvectors.

\subsection{\textit{Fastest and Slowest Growth Clusters}}\label{CDgrowth}

For future analysis (\autoref{sigfeatures}), it is useful to define the clusters that contain the counties with the fastest and slowest growth. After the clusters are produced with Algorithm~\ref{alg:Algo1}, for every community $k$, we calculate the average exponential growth rates of the counties in that community. This is done by fitting the total number of cases of each county $i$ on day $t$, $x_{i,t}$, to $x_{i,t} = x_{i,0}(1+r_i)^t + \varepsilon_{i,t}$ through nonlinear least squares and obtaining the approximated growth rate $r_i$ for county $i$. 
Then, we compare the average fitted growth rate $\hat{r}_k = 1/|\hat{G}_k| \sum_{i\in \hat{G}_k} r_i$ and standard error for clusters $k = 1,...,K$. The fastest growth cluster is defined as $\argmax_{k} \hat{r}_k$ and the slowest growth cluster is defined as $\argmin_{k} \hat{r}_k$. 


\subsection{\textit{Data}} \label{CDdata}
We use the COVID-19 (2019-nCoV) Data Repository by the Johns Hopkins Center for Systems Science and Engineering (CSSE) that contains data on the number of confirmed cases and deaths in the United States and around the world, broken down by counties in the U.S. The public database is updated daily and the virtual dashboard is also used widely around the world. Data sources of the database include the World Health Organization (WHO), US Center for Disease Control (CDC), BNO News, WorldoMeters, and 1point3acres. We take all counties that have $12$ or more cumulative cases in the time frame of 01/22/2020 to 04/17/2020. We treat the day a county reaches $12$ or more confirmed cases as day one and then discard all counties that have a time series of fewer than 14 days after processing. This way we shift each county to a similar starting point in terms number of cases and a long enough period to do a meaningful analysis with. We also remove unassigned cases and U.S. territories, which ultimately results in a total of 950 counties. As mentioned before, we use $w_{i,t} = \log({x_{i,t}})$ to represent the logarithmic cumulative cases for county $i$ on day $t$. 

We also repeat the community detection process with more recent data from 05/10/2020, when many states started to reopen, to 07/10/2020. The bulk of this part of the study concentrates on the beginning phase of the pandemic given that health and government intervention to minimize the number of future cases should be executed as early as possible. However, we compare the resulting communities with more recent data that captures the second phase of the pandemic in the U.S. States experienced a significant drop in cases when lockdown was enforced and businesses were closed but as they began to reopen, the number of cases saw an uptick once again. Since this second phase comes months after the initial outbreak, there may be meaningful differences worthy of analysis.

\begin{table*}\centering
\footnotesize 
\begin{tabular}{@{}lllclllclll@{}}\toprule 
& \multicolumn{3}{c}{Group 1} & \multicolumn{3}{c}{Group 2} \\
\midrule
	$Model$& No. of Counties & Growth Rate & SE & No. of Counties & Growth Rate & SE\\ \midrule
$\boldsymbol{R}$ &467 & 0.1589&0.0020&483	& 0.1704 &0.0019\\
$\boldsymbol{A}_1$ & 462& 0.1583 &0.0020&488 & 0.1677&0.0019\\
$\boldsymbol{A}_2$ & 470& 0.1605&0.0020&470 &0.1664&0.0020\\
$\boldsymbol{R}$, second phase & 487 &0.0207& 0.0005&463 &0.0233&0.0005\\
\bottomrule
\end{tabular}
\caption{\footnotesize Average growth rates and the number of counties in each cluster for $K=2$. Model $\boldsymbol{R}$ corresponds to Algorithm~\ref{alg:Algo1} where we use the sample correlation matrix. Model $\boldsymbol{A}_1$ corresponds to Algorithm~\ref{alg:Algo4} where we use the $k$-nearest neighbors graph ($k=7$). Model $\boldsymbol{A}_2$ corresponds to Algorithm~\ref{alg:Algo4} where we use the $\epsilon$-neighborhood graph ($\epsilon = 0.007$). Groups 1 and 2 are the obtained partitions $\hat{G}_1$ and $\hat{G}_2$, respectively. Growth Rate is the approximated exponential growth rate, calculated as in \autoref{CDgrowth}. Presented are the averages of these growth rates and their associated SEs for the counties in two groups, clustered by different methods.  $\boldsymbol{R}$, second phase is for the clusters obtained for the period 05/10/2020 - 07/10/2020. }
\label{Tab:overview}
\end{table*}

\vspace{-0.3in}
\begin{table*}\centering
\begin{tabular}{@{}lllclllclll@{}}\toprule 
Group & 1& 2 \\ \midrule
1&96.1$\%$&94.0$\%$\\
2&94.0$\%$&96.8$\%$\\
\bottomrule
\end{tabular}
\caption{\footnotesize $\boldsymbol{R}$ average block correlations $K=2$.}
\label{Tab:block}
\end{table*}

\subsection{\textit{Results and Discussion}} \label{CDresults}

\begin{wrapfigure}{R}{0.4\textwidth}
  \centering
  \includegraphics[width=0.4\textwidth]{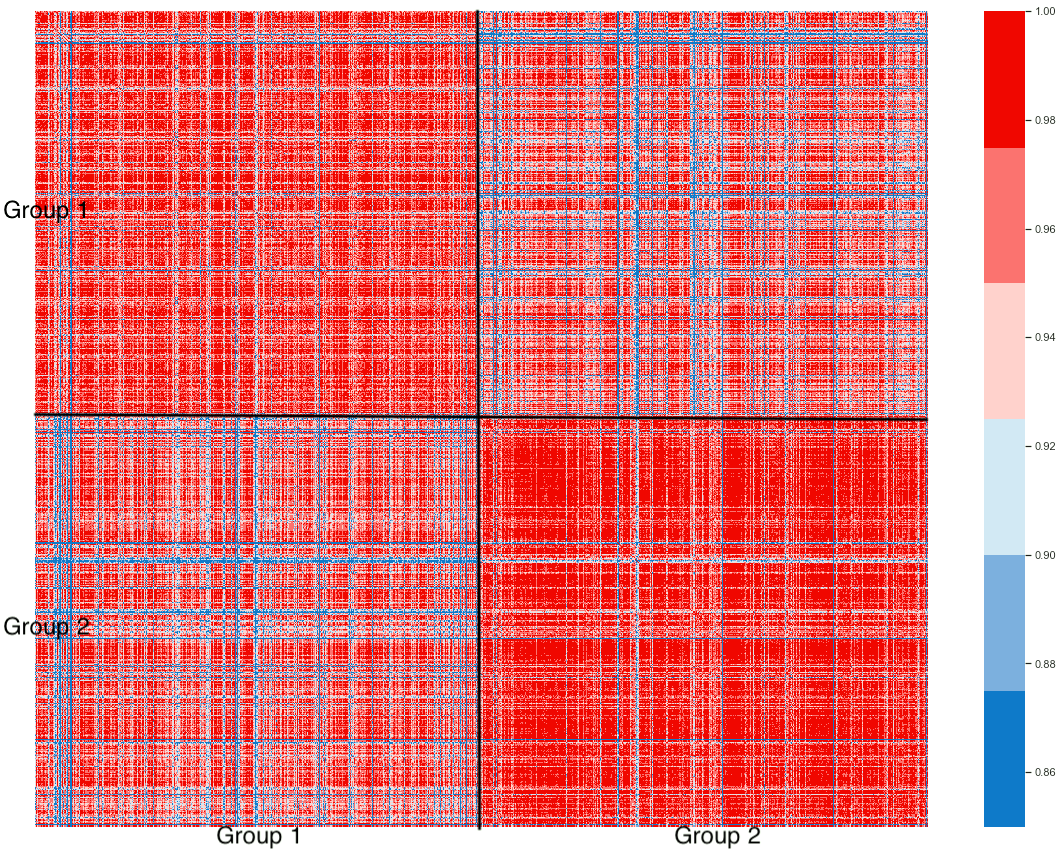}
  \vspace{-0.3in}
   \caption{\footnotesize $\boldsymbol{R}$ Heatmap of block correlations $K=2$. Model $\boldsymbol{R}$ corresponds to Algorithm~\ref{alg:Algo1} where we use the sample correlation matrix. Groups 1 and 2 are the obtained partitions $\hat{G}_1ß$ and $\hat{G}_2$, respectively.}\label{fig:heatmap}
   \vspace{-0.2in}
\end{wrapfigure}

\begin{figure}
    \centering
    \includegraphics[width = 17cm, height = 6cm,valign=t]{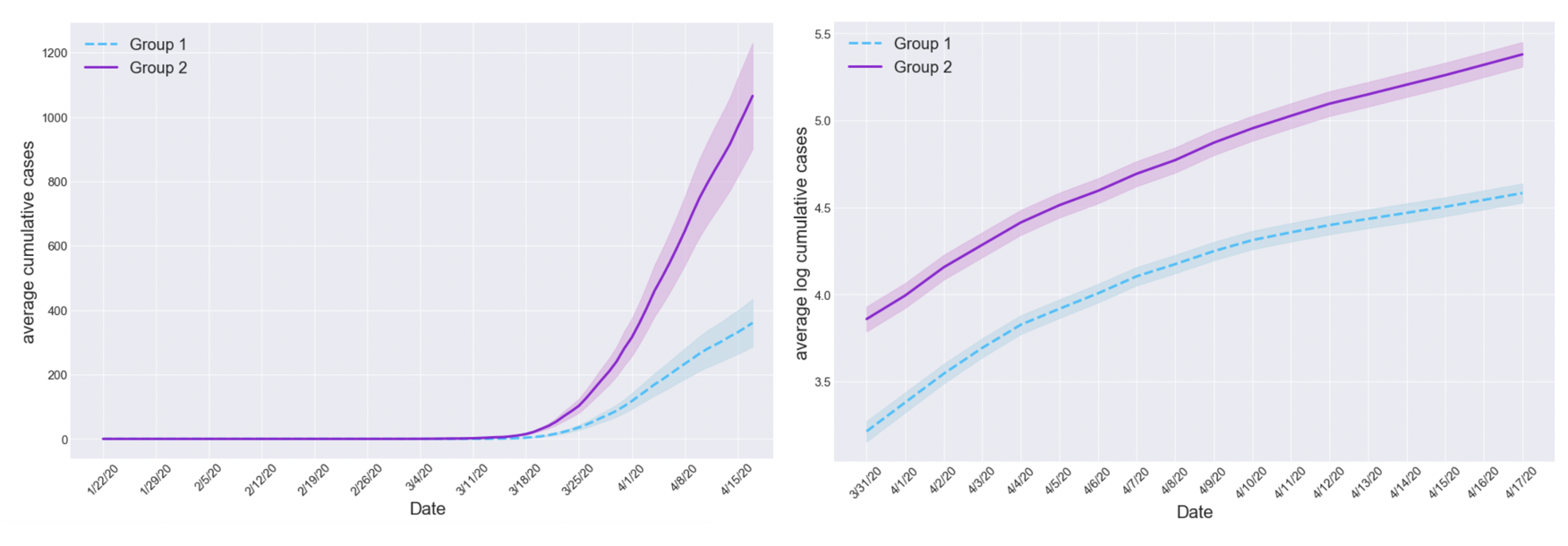}
    \vspace{-0.2in}
    \caption{\footnotesize The left panel represents the average cumulative number of cases of the initial phase 01/22/2020 - 04/17/2020 with one standard error bands for the clusters of $\boldsymbol{R}$,  $K=2$. The right panel is the average log cumulative number of cases of $\boldsymbol{R}$, $K=2$. The x-axis is in calendar time, which does not account for heterogeneous starting times of the outbreak in each county.}
    \label{fig:casesdates}
    \centering
    \includegraphics[width = 17cm, height = 6cm,valign=t]{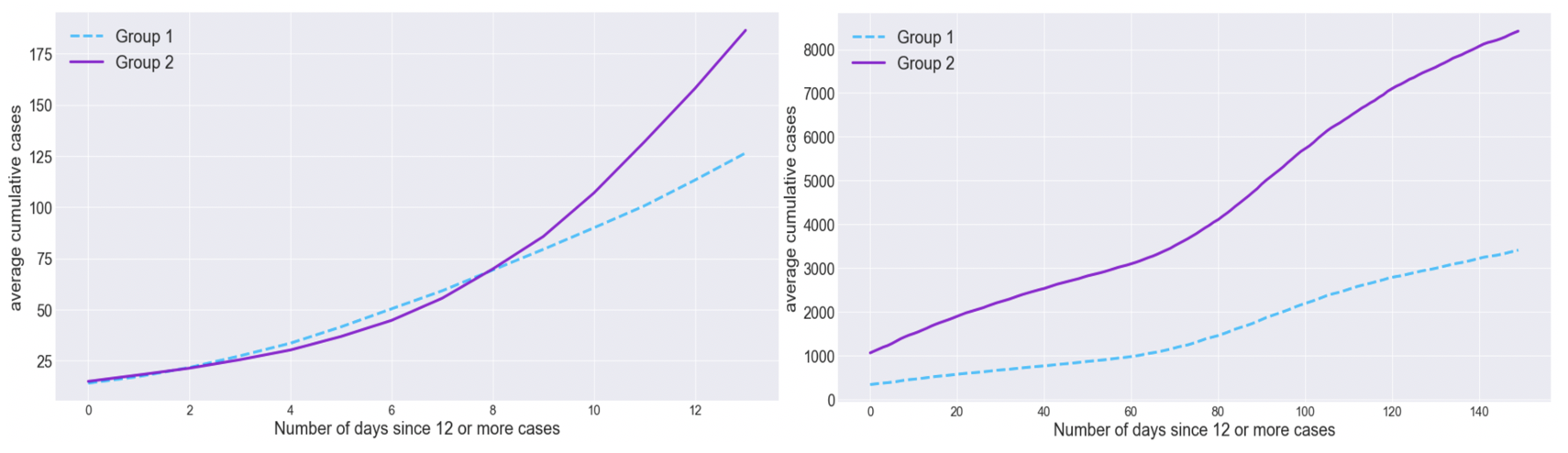}
    \vspace{-0.15in}
    \caption{\footnotesize The left panel represents average cumulative number of cases of the initial phase 01/22/2020 - 04/17/2020, starting from the first day of at least 12 days for the clusters of $\boldsymbol{R}$,  $K=2$. The right panel is the average cumulative number of cases of the period 04/18/2020 - 09/03/2020, the time frame after the initial phase used in community detection.  The x-axis here accounts for the heterogeneity of the outbreak of COVID-19 in each county.}
    \label{fig:casessince12}
      \includegraphics[width = 15cm,height = 6cm,valign=t]{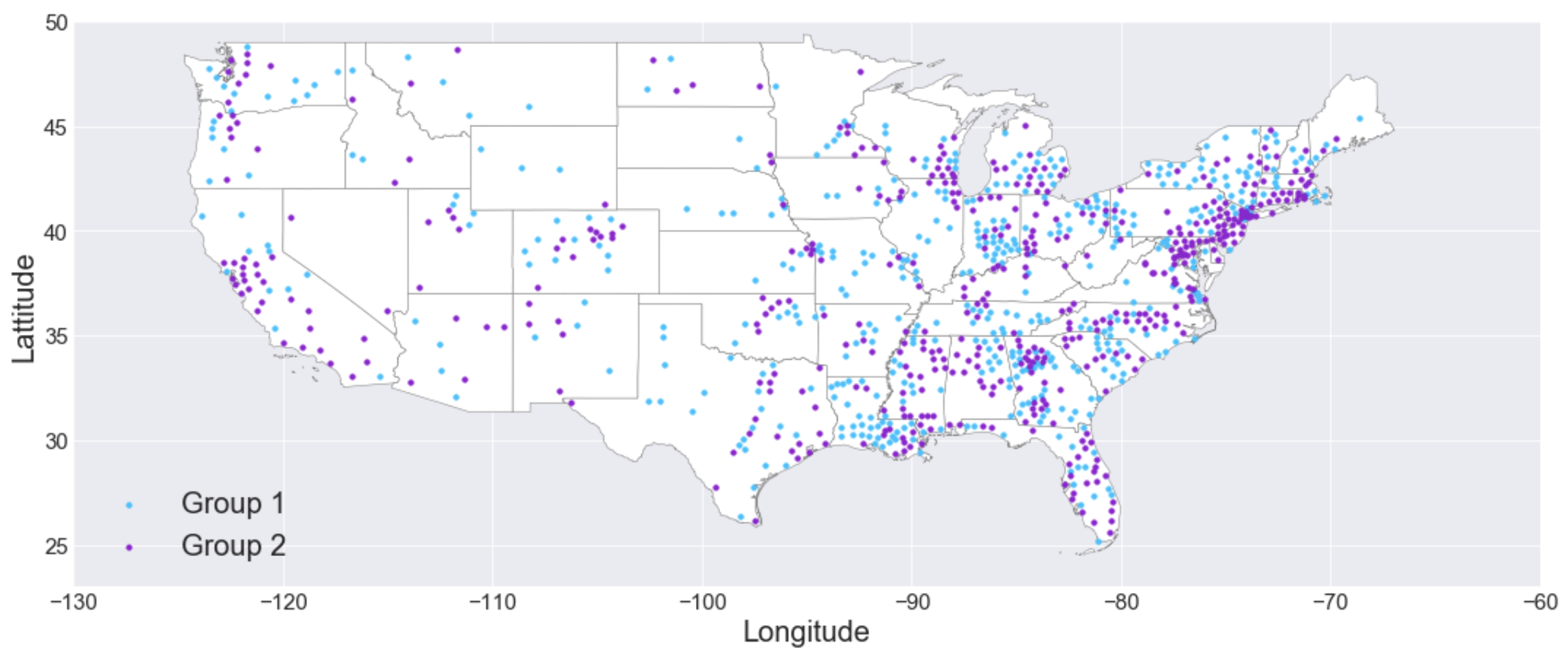}
    \vspace{-0.2in}
       \caption{\footnotesize Clusters for model $\boldsymbol{R}$ of the initial phase 01/22/2020 - 04/17/2020. Model $\boldsymbol{R}$ corresponds to Algorithm~\ref{alg:Algo1} where we use the sample correlation matrix. Groups 1 and 2 are the obtained partitions $\hat{G}_1$ and $\hat{G}_2$, respectively.}\label{fig:corrclusters}
\end{figure}

We can see from Table \ref{Tab:overview} that for the clusters obtained by Algorithm~\ref{alg:Algo1} ($\boldsymbol{R}$), the difference between the growth rates of Group 1 and Group 2 is the largest. This is validated by the discernible difference in trajectories, shown in Figures \ref{fig:casesdates} and \ref{fig:casessince12}. The standard error bands in Figure \ref{fig:casesdates} underscores that the two groups become more distinct in their growth trajectory as time goes on: the overlap between the bands of the two groups decreases over time. For $\boldsymbol{A}_1$ and $\boldsymbol{A}_2$, the growth rates are much closer together between the two communities. Furthermore, the right panel of Figure 6 is a plot of the average cases for the period after community detection was performed: 04/17/2020 - 09/03/2020. Evidently, the separation between the two groups becomes much more distinct as time goes on (much larger number of cases). As for community detection of the subsequent phase of the pandemic in the U.S. (from 05/10/2020 - 07/10/2020), the last row of Table \ref{Tab:overview} again shows a larger average growth rate for Group 2, albeit much smaller in magnitude since cases increased at a slower rate once the country learned how to deal with the pandemic. 

Table \ref{Tab:block} contains information on the average intra- and inter-group correlations, a sample reflection of $\boldsymbol{B}$. Evidently, the intra-community correlations are higher than the inter-community correlations. Group 1's intra-correlation of 96.1\% and Group 2's 96.8\% are greater than 94.0\%, the inter-group correlation between the two groups. As we only took counties with significant outbreaks as of 04/17/2020, it is logical to observe high correlations across the board. 
These results are also mirrored in Figure \ref{fig:heatmap}, a heatmap of the block correlations. 

Some notable counties that are partitioned to Group 2, the fast growth community, include Los Angeles, CA; San Francisco, CA; District of Columbia; DeKalb, GA; Fulton, GA; Miami-Dade, FL; Cook, IL; Jefferson, LA; Suffolk, MA; Bergen, NJ; New York, NY; Westchester, NY; and King, WA, all large epicenters. Figure \ref{fig:corrclusters} is a geographical visualization of the communities.

\begin{figure}
    \centering
    \includegraphics[height = 6cm, width = 17cm,valign=t]{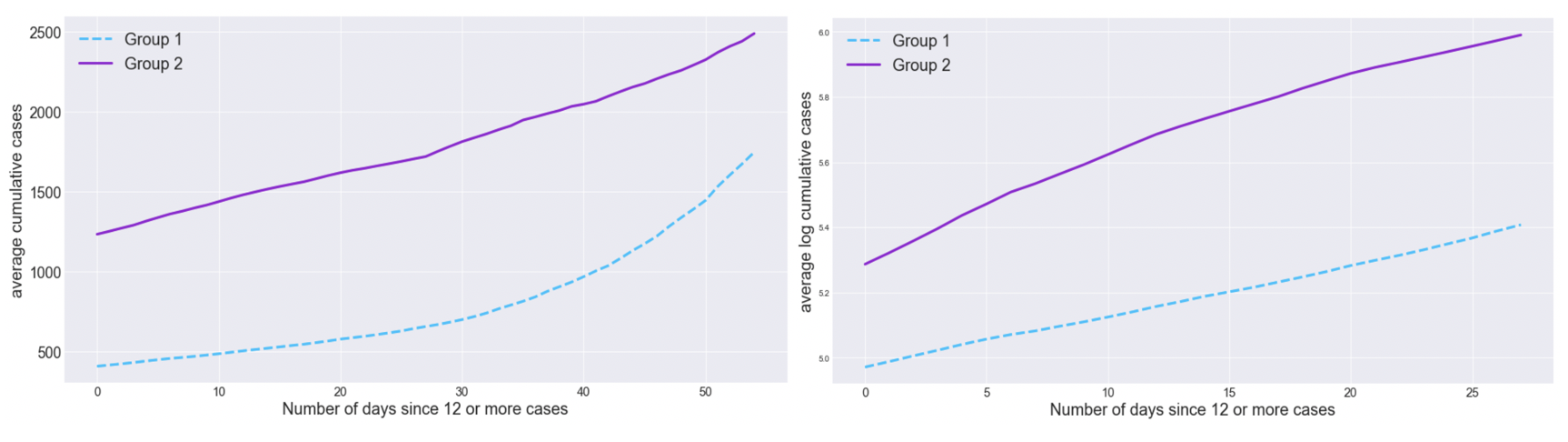}
     \vspace{-0.2in}
    \caption{\footnotesize Growth curves of clusters obtained from community detection on data from 05/10/2020 - 07/10/2020 (recent phase). Plots are the same as those of Figure \ref{fig:casessince12}.}
    \vspace{-0.2in}
    \label{fig:recent-cases12}
\end{figure}

In addition, Figure \ref{fig:recent-cases12} shows the same plots as those in Figure \ref{fig:casessince12} but for a later phase. The curves are clearly much flatter in both groups, which is likely due to the increase in the number of cases plateauing in many counties. Furthermore, the distinction between the curves of Group 1 and 2 is also considerably bigger than those of the earlier data. This can be explained by the confluence of additional factors that separate each county's experience with the virus, including the nature of local government intervention, degree, and timing of re-openings, travel restrictions, etc. 

\begin{center} \section{EXTRACTING SIGNIFICANT DEMOGRAPHIC FEATURES} \end{center} \label{sigfeatures}
An important and subsequent question that arises once the communities are obtained is what underlying factors play a role in which growth cluster a county belongs to. Since the growth of COVID-19 cases is also related to static, inherent factors that aren't a consequence of the disease, we examine a variety of county demographic variables and how they differ among communities. In order to select the variables that are most statistically significant, or are most relevant to the community assignment of a county, we perform independent two-sample t-tests on the fastest and slowest growth groups (\autoref{CDgrowth}) with respect to various demographic variables. The null and alternative hypotheses for this t-test for the $d$-th feature are as follows: 
\begin{equation}\label{eq:hyp}
    H_0: \mu_{d,1} = \mu_{d,2}, \: vs. \: H_a: \mu_{d,1} \neq \mu_{d,2},
\end{equation}
where $\mu_{d,1}$ is the mean value of the $d$-th feature of cluster $1$ and $\mu_{d,2}$ is the mean value of the $d$-th feature of cluster $2$. We then compute the two-sample test statistic with pooled estimate of the variance.  
After finding the p-values, we rank the features from lowest p-value (most significantly different between the two groups) to highest (least significantly different between the two groups).

Furthermore, we repeat Algorithm~\ref{alg:Algo1} for $K = 3,4,5$, select the 'fastest' and 'slowest' growth clusters in each case, and carry out the independent two-sample t-tests as described above for the same demographic features. This sensitivity analysis tests whether the demographic variables that are the most significantly different between the two groups are consistent when we have a larger number of communities.  Ultimately, we present the most statistically significant demographic features.

\subsection{\textit{Data}} \label{SFdata}

For this section, we use data from Data Planet, a social science research database that compiles 12.6 billion U.S. and international datasets from over 80 sources. For our purposes, we look at the 2017 American Community Survey (ACS), the largest household survey in the U.S., conducted by the U.S. Census Bureau. We select 17 relevant features on a county-level, which are displayed and summarized in Table~\ref{Tab:corr}. Note that not all $950$ counties from Johns Hopkins CCSE data that were used in \autoref{CDdata} is available on Data Planet, thus the analysis is done on $633$ counties for this section. Now, we are left with $301$ counties in Group 1 and $332$ counties in Group 2, which is still a close split like that of $\boldsymbol{R}$ seen in Table~\ref{Tab:overview}.

\begin{figure}
    \centering
    \includegraphics[scale=0.52,valign=t]{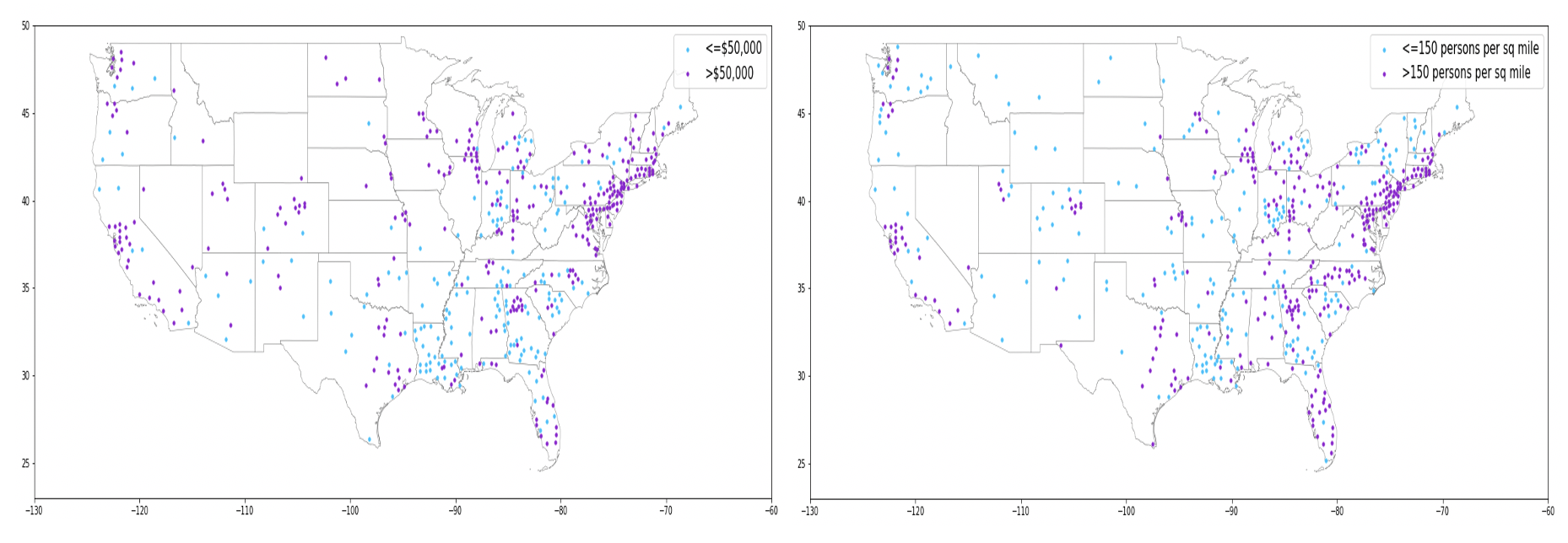}
    \caption{\footnotesize 
      The left panel is a geographical representation of counties according to median household income. Blue dots are counties with less than \$50,000 median annual household income and purple dots are counties with more than \$50,000 median annual household income. The right panel is a geographical representation of counties according to population density. Blue dots are counties with less than 150 persons per sq mile and purple dots counties with more than 150 persons per sq mile. }
    \label{fig:pdandmi}
    \vspace{-0.3cm}
\end{figure}

\makeatletter
\setlength{\@fptop}{0pt}
\makeatother

\begin{table*}[h!]
\centering
\footnotesize
\resizebox{\columnwidth}{4.5cm}{
\begin{tabular}{@{}lllclllclll@{}}\toprule 
& \multicolumn{3}{c}{Group 1} & \multicolumn{3}{c}{Group 2}\\
\cmidrule{2-3} \cmidrule{4-5} \cmidrule{6-7}
	$Feature$& Mean & Median & Std Dev & Mean & Median & Std Dev\\ \midrule
Population Density & 275.775& 159.260& 394.059 & 913.182 & 289.610 & 3659.76\\
Median Household Income & 54431.6 &52651.0 &13386.3& 58814.6& 56074.0& 16737.2\\
$\%$ Poverty & 14.0656&13.3000&5.71511&13.4712&12.5000&5.83384\\
$\%$ 1-person households &27.1330&27.7185&4.21220&27.0233&27.1978&4.54690\\
$\%$ 5 or more person households &8.91332& 8.41406& 2.96792& 9.38467& 8.71688& 3.27316\\
$\%$ households w 60 y/o and older &39.3294& 39.1802& 6.85032& 38.7538& 38.6947& 6.03233\\
$\%$ w low access to stores &21.8723&21.3800&9.66507&20.8704&21.2500&9.81211\\
$\%$ low income w low access to stores &7.48273&6.85500&4.48466&6.60216&5.69000&4.52299\\
$\%$ households w low access to stores &2.69416&2.32000&1.63712&2.24196&1.92000&1.50078\\
25 y/o and older w Bachelor's /1,000 &110.885&106.164&40.6371&122.654&118.332&43.6276\\
$\%$ White &80.8599&85.6959&15.0719&75.7593&79.6171&16.7522\\
$\%$ Black & 11.3861&5.03010&14.6140&13.4737&7.90300&15.3919\\
$\%$ Asian &1.96190&1.22070&1.89220&3.67570&1.87900&5.28820\\
No of bars&29.3313&16.0000&39.0755&55.2143&23.5000&96.9646\\
No of grocery stores&42.5564&23.0000&67.0810&117.321&39.0000&16.5490\\
No of restaurants&13.1345&8.00000&13.1383&14.9219&9.00000&16.5490\\
$\%$ take public transportation&0.41130&0.19870&0.76870&1.24690&0.32130&6.14170\\
\bottomrule
\end{tabular}}
\caption{\footnotesize $\boldsymbol{R}$ clusters' mean and median values for selected features for each community $K=2$. Model $\boldsymbol{R}$ corresponds to Algorithm~\ref{alg:Algo1} where we use the sample correlation matrix. Group 1 and 2 are the obtained partitions $\hat{G}_1$ and $\hat{G}_2$, respectively. Population Density is the number of people per sq mile; median household income is in US dollars; \% Poverty is the poverty rate: \% 1-person households is the percentage of one-person households; \% 5 or more person households is the percentage of five or more person households; \% households w 60 y/o and older is the percentage of households that have one or more members who are 60 years old or older; low access to stores is defined as living more than one mile (urban areas) or 10 miles (rural areas) from the nearest supermarket, supercenter, or large grocery store; /1,000 is per 1,000 persons; \% take public transportation is the percentage of all persons who work in a county and take public transportation to work every day. All feature information is as of 2017.}

\label{Tab:corr}
\end{table*}

\begin{figure*}[h!]
    \centering
    \includegraphics[scale=0.35,valign=t]{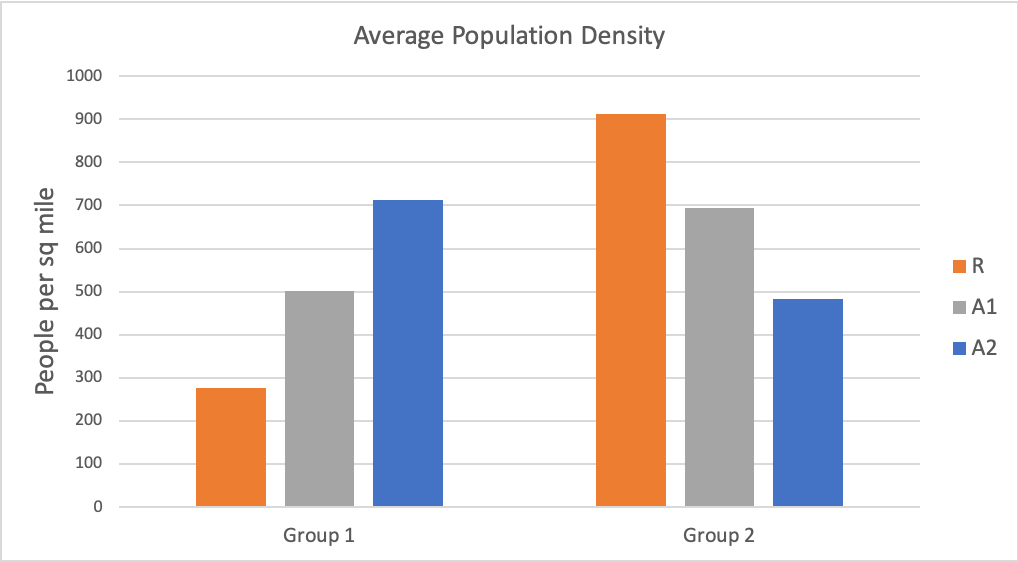}
    \includegraphics[scale=0.35,valign=t]{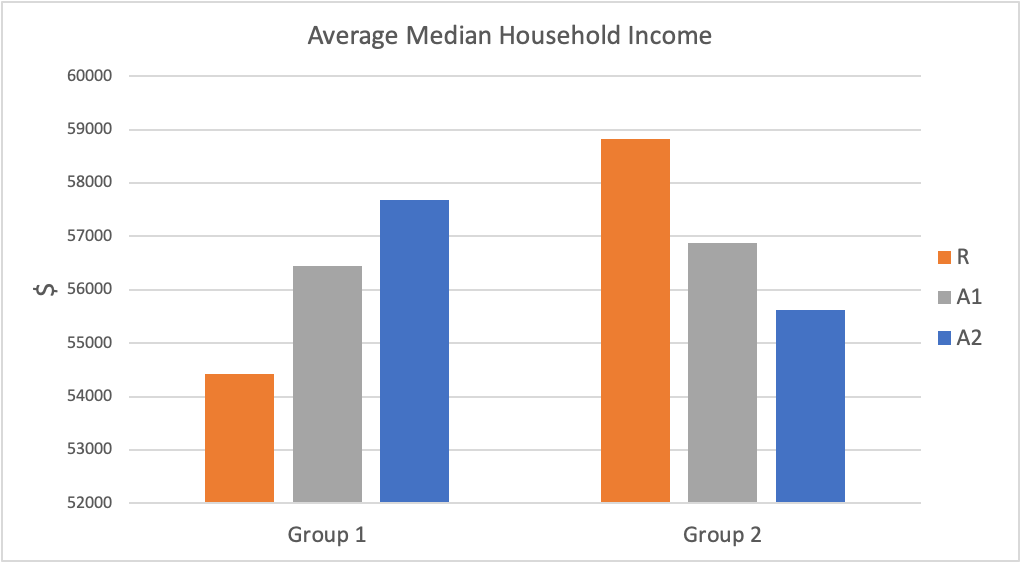}
    \includegraphics[scale=0.35,valign=t]{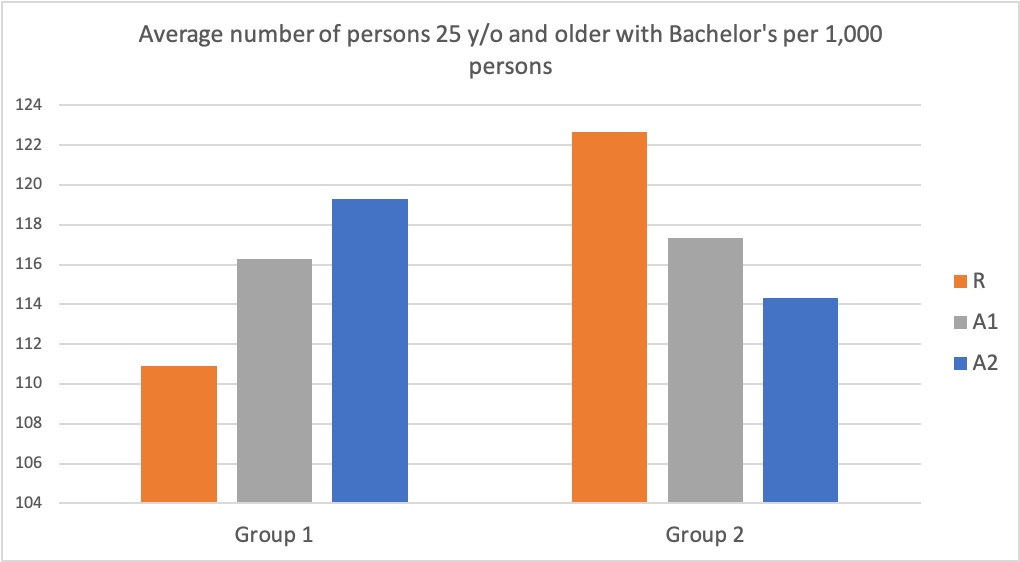}
   \includegraphics[scale=0.35,valign=t]{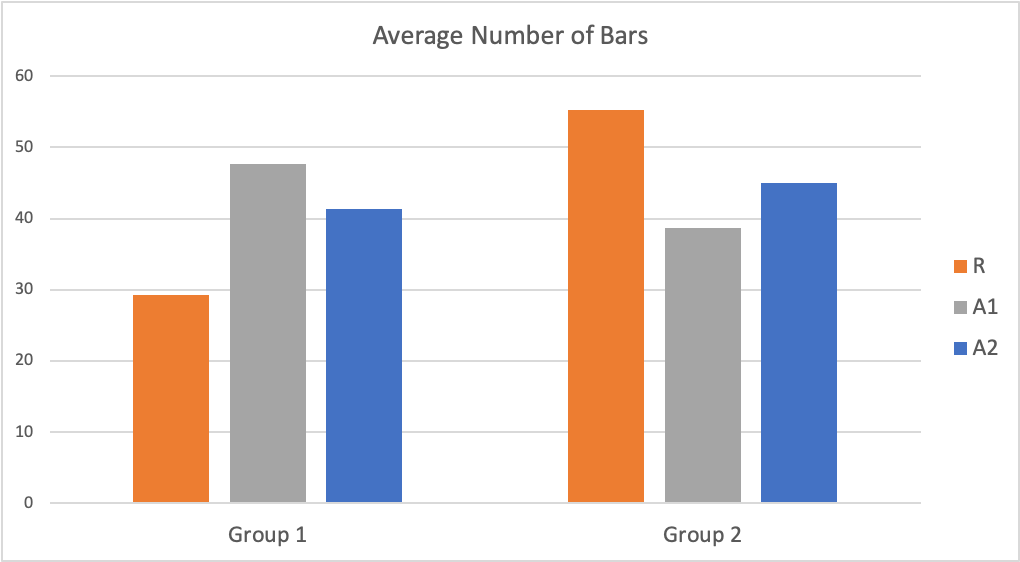}
    \includegraphics[scale=0.35,valign=t]{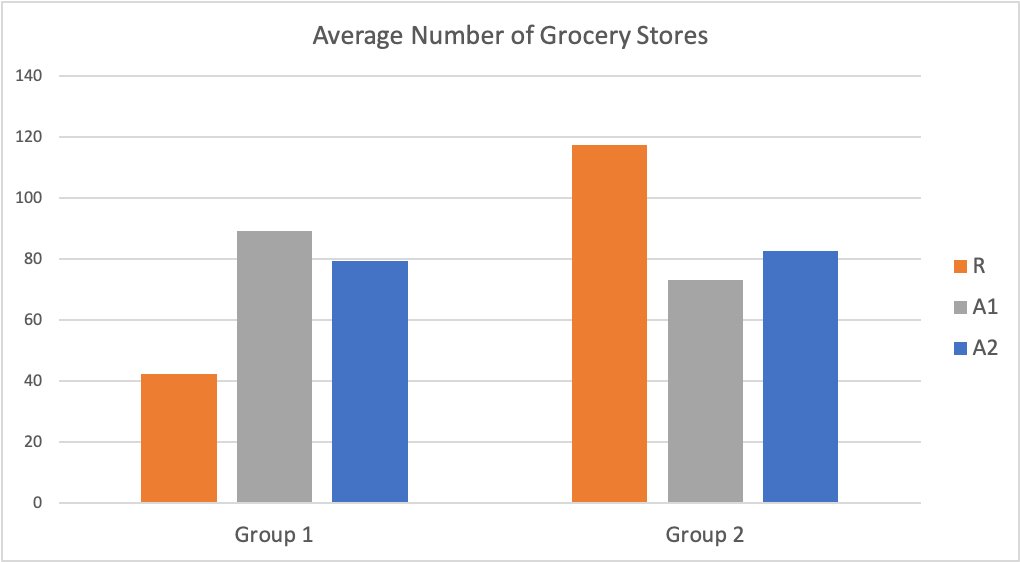}
    \includegraphics[scale=0.35,valign=t]{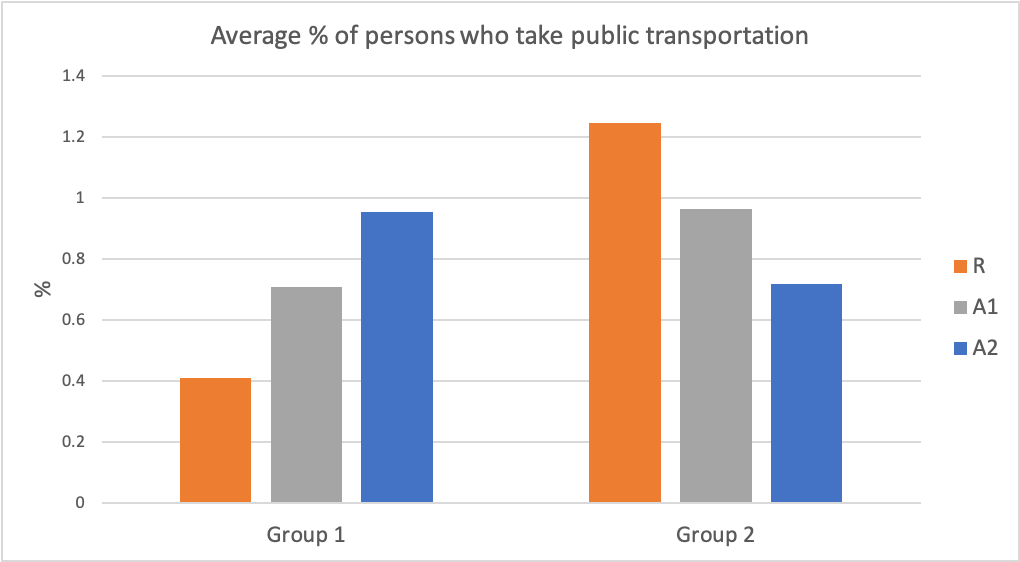}
    \caption{\footnotesize Comparisons on the averages of a few features of the  2-clustered counties. The orange bars are model $\boldsymbol{R}$, the gray bars are model $\boldsymbol{A}_1$, and the blue bars are model $\boldsymbol{A}_2$.} 
    \label{fig:feature plots}
\end{figure*}

\subsection{\textit{Results and Discussion}} \label{SFresults}

\begin{table*}[h!] \centering
\footnotesize
\begin{tabular}{@{}lllclllclll@{}}\toprule 
	$Feature$& P-Value&$Feature$& P-Value\\ \midrule
\textbf{\% Asian}&1.26E-11 &\textbf{No of grocery stores}& 2.85E-08\\
\textbf{No of grocery stores}&5.83E-11 &\textbf{\% low income w low access to store} &6.38E-07\\
\textbf{No of bars}&2.95E-06&\textbf{Median Household Income} & 8.41E-06\\
\textbf{\% White} &3.04E-06&\textbf{\% Poverty} & 2.01E-04\\
\textbf{Median Household Income}&1.44E-05&\textbf{\% White} & 0.01967\\
\% households w low access to stores&1.60E-05&\textbf{Population Density} & 0.04027\\
\textbf{25 y/o and older w Bachelor's /1,000}&2.80E-05&\textbf{\% 1-person households}& 0.07537\\
\textbf{Population Density}&2.39E-04&\% households w low access to stores & 0.10640\\
\% low income w low access to store&0.00303&\% 5 or more persons households& 0.10711\\
\% take public transportation&0.00436&\% Black & 0.12222\\
\% 5 or more persons households&0.02324&\% take public transportation &0.13245\\
\% Black&0.04672&\% of households w 60 y/o and older & 0.35344\\
\% Poverty&0.10603&\% Asian & 0.47698\\
No of restaurants&0.11432&25 y/o and older w Bachelor's /1,000& 0.52498\\
\% w low access to stores&0.13698&\% w low access to stores & 0.62138\\
\% households w 60 y/o and older&0.18014&No of restaurants &0.89977\\
\% 1-person households&0.63999&No of bars&0.90224\\\midrule
\end{tabular}
\caption{\footnotesize Left table is $\boldsymbol{R}$ clusters' p-values for independent two-sample t-tests for selected features between Group 1 and Group 2 sorted from smallest to largest p-value. Right table is recent data (05/10/2020 - 07/10/2020) $\boldsymbol{R}$ clusters' p-values. The features in bold are the ones that are selected as significant features for further analysis in \autoref{predictionsd}.}
\label{Tab:pvaluesR}
\end{table*}

\begin{table*}[h!]
\centering 
\footnotesize
\begin{tabular}{@{}lllclllclll@{}}\midrule
	$Feature$& P-Value&$Feature$ & P-Value\\ \midrule
\% w low access to stores &0.02683&\% White &	0.09115\\
\% low income w low access to store &	0.08490&\% Poverty &	0.09423\\
No of bars &0.12402&\% households w low access to stores& 	0.12762\\
No of restaurants &	0.13868&Median Household Income &	0.16226\\
No of grocery stores &	0.18422&Population Density &	0.17828\\
\% Black &0.21146&25 y/o and older w Bachelor's /1,000 &0.32282\\
\% of households w 60 y/o and older &0.22823&\% low income w low access to stores& 	0.36051\\
Population Density &0.27201&No of restaurants &0.39537\\
\% 5 or more persons households& 0.34981&\% take public transportation &	0.50781\\
\% take public transportation&0.38932&\% Asian &	0.51744\\
\% 1-person households& 0.52795&\% 1-person households & 0.41733\\
\% households w low access to stores &0.46620&\% Black &	0.56017\\
\% Poverty &	0.63207&\% households w 60 y/o and older &0.57800\\
\% White &	0.47153&No of grocery stores &	0.72219\\
Median Household Income &0.67262&\% w low access to stores& 0.78404\\
25 y/o and older w Bachelor's /1,000&	0.71546&\% 5 or more persons households&	0.91679\\
\% Asian &	0.92895&No of bars &0.93340\\
	\bottomrule
\end{tabular}
\caption{\footnotesize Left panel is $\boldsymbol{A}_1$ clusters' p-values for independent two-sample t-tests for selected features between Group 1 and Group 2 sorted from smallest to largest p-value. The right panel is $\boldsymbol{A}_2$ clusters' p-values for two-sample t-tests for selected features between Group 1 and Group 2 sorted from smallest to largest p-value.}
\label{Tab:pvaluesA}
\end{table*}
It is evident from Table \ref{Tab:corr} that community detection with $\boldsymbol{R}$ results in Group 2 (fast growth) containing counties with the highest mean and median population density by far. The mean and median household income are also higher for counties in Group 2. The mean number of persons 25 years old or older with Bachelor's is noticeably greater for Group 2, which can often coincide with more urban areas that are more densely populated. However, it can also be related to the number of universities in a particular area, as a higher number would exacerbate the spread of COVID-19. The number of bars and grocery stores are also starkly different among the two groups. Moreover, the percentage of people who take public transportation to work is around three times greater for Group 2 than Group 1. These observations do not hold for $\boldsymbol{A}_1$ and $\boldsymbol{A}_2$ (see Supplementary Material for Tables \ref{Tab:A1} and \ref{Tab:A2}) and in fact, the differences in mean and medians between the two groups are significantly smaller for almost all features. For instance, the differences between population density means for the two groups of $\boldsymbol{A}_1$ and $\boldsymbol{A}_2$ are 191.609 and 228.966 people per sq mile, respectively - much smaller differences than that of $\boldsymbol{R}$'s clusters: 637.407. Furthermore, there isn't consistency in these trends for most features; for example, Group 2 of $\boldsymbol{A}_1$ has the higher population density and median income averages but also has the lower average number of bars, grocery stores, and restaurants. 

On the other hand, unlike what one would expect in terms of the relationship between the number of one-person households and the spread of COVID-19, there is not much differentiation between the number of people in a household. Figure \ref{fig:feature plots} displays the individual bar plots of average values for six features for $\boldsymbol{R}$, $\boldsymbol{A}_1$, and $\boldsymbol{A}_2$. It is evident that the orange bars of $\boldsymbol{R}$'s clusters are plainly contrasted between Group 1 and 2, unlike those of $\boldsymbol{A}_1$ and $\boldsymbol{A}_2$'s clusters. It is very likely that too much information was lost in $\boldsymbol{A}_1$ and $\boldsymbol{A}_2$ during the truncation process.

Tables \ref{Tab:pvaluesR} and \ref{Tab:pvaluesA} contain the p-values for all of the 17 features. Evidently, the clusters formed with $\boldsymbol{A}_1$ and $\boldsymbol{A}_2$ do not contain as significant differences in terms of demographic variables as they have much higher p-values across the board. As for $\boldsymbol{R}$, the variables have a much greater explanatory power. The number of grocery stores and the number of bars have much lower p-values than that of the number of restaurants. Also, as expected, the median household income is among the features with lower p-values; however, population density does not appear as significant as expected. After conducting the same two-sided t-tests for $K = 3,4,5$ on the two extreme groups, the seven statistically significant features found are as follows: population density, median income, number of persons who are 25 years and older with Bachelor’s per 1,000 persons, percentage of the White population, percentage of the Asian population, number of bars, and number of grocery stores. These seven features are consistently ranked in the top eight features for each $K = 2,3,4,5$ based on p-values. These values form the demographic vector $\boldsymbol{d_i}$ for each county $i$. The variables for bars and grocery stores underscore the ease of transmission in locations with greater numbers of public gathering spots, a characteristic evident in cities like New York City where most people choose to convene at bars without much social distancing (before stricter lockdowns took place). 

After finding the growth communities and conducting t-tests to ascertain the significant features for the latter phase of the pandemic in the U.S. (05/10/2020 - 07/10/2020), the features with the lowest p-values diverge from those of earlier data, as presented in the right table of Table \ref{Tab:pvaluesR}. Population density and median income are still among the most meaningful, with income being more significant, but variables such as the number of grocery stores, percentage of people with low access to stores, and the percentage living in poverty have become much more significant. This suggests that at later stages of the pandemic, poverty and other income-related measures become more indicative and responsible for the differences in case growth among counties. Thus, the seven features for $\boldsymbol{d_i}$ for this latter phase 
are the top seven variables in Table \ref{Tab:pvaluesR}: number of grocery stores, \% low income with low access to stores, median household income, \% poverty, \% white, population density, and \% 1-person households.

\begin{center} \section{PREDICTION WITH SOCIAL DISTANCING DATA}\label{predictionsd} \end{center} 
The final section of our COVID-19 methodology is to predict a county's growth trajectory a few days into the future. We propose a prediction methodology with the objective that given a new county, the new county's key demographic features, and social distancing measures, we implement an algorithm that projects the new county's future growth.

Before going in-depth on the prediction models, it's necessary to first define some important variables. 
Let $l$ be the number of the days forward to be projected for a new county.  To build such a predictive model, let $y_{i,t+l} = \log(x_{i,t+l}) - \log(x_{i,t})$ be county $i$'s $l$-day forward log-growth rate, which is close to the growth rate $\frac{x_{i,t+l} - x_{i,t}}{x_{i,t}}$ by Taylor's expansion and numerical verification, for $t = 1,...,T_i$.  Here,  $T_i+l$ is the total number of days where county $i$ has 12 or more cases.  Recall the obtained partitions from Algorithm~\ref{alg:Algo1} (set of indices of counties that belong to group $k$): $\hat{G}_k = \{i | \hat{Z}_i = k, i = 1,...,n\}$, where $\boldsymbol{\hat{Z}} \in \mathbb{R}^n$ is the recovered community label vector.      For a community $k$, and a county $i \in \hat{G}_k$, let $\boldsymbol{d}_i \in \mathbb{R}^q$ be county $i$'s significant feature vectors obtained from \autoref{sigfeatures},  $\boldsymbol{S}_i = [\boldsymbol{s}_1^i,\boldsymbol{s}_2^i,...,\boldsymbol{s}_{T_i}^i]^T \in \mathbb{R}^{T_i \times 3}$ be county $i$'s three social distancing time series matrix (see \autoref{preddata} for details about this data) and $\by_i = [y_{i,1+l}, \cdots, y_{i, T_i+l}]^T \in \mathbb{R}^{T_i}$ be its $l-$day forward log case difference.  Note that each row of $\boldsymbol{S}_i$, $\boldsymbol{s}_t^i \in \mathbb{R}^3$, has three different social distancing metrics at time $t$.  

In summary, we have data $\{\bS_i, \by_i: i \in \hat{G}_k\}$ for training an $l$-day ahead predictive model for the $k^{th}$ community.

\subsection{\textit{Long Short-Term Memory (LSTM)}} \label{LSTM}
To enhance the effectiveness of the model, we take advantage of a special type of recurrent neural network (RNN): long short-term memory (LSTM) networks, which are designed for time-series forecasting. Unlike feedforward neural networks (FNNs), RNNs produce an output that depends on a \say{hidden} state vector that contains information based on prior inputs and outputs. LSTMs builds on a simple, vanilla RNN to include a forget gate, input gate, and output gate for each module. Hence, it is able to \say{remember} information for longer time periods (lags). The output for an LSTM module at time $t$ is as follows:
\begin{equation}\label{eq:lstm}
    h_t = o_t \tanh(C_t).
\end{equation}

The components of $h_t$ are broken down as follows: $f_t = \sigma(W_f[h_{t-1},x_t]+b_f),$ is the forget gate output and $W_f$ and $b_f$ are its weights and biases, respectively. $  i_t = \sigma(W_i[h_{t-1},x_t]+b_i),$ is its input gate output and $W_i$ and $b_i$ are its weights and biases, respectively. The cell state vector then gets updated by forgetting the previous memory through the forget gate and adding new memory through the input gate: $C_t = f_tC_{t-1} + i_t\Tilde{C}_t,$ where $ \Tilde{C}_t = \tanh(W_C[h_{t-1},x_t]+b_C)$. Subsequently, the output gate $o_t = \sigma(W_o[h_{t-1},x_t]+b_o)$ and $W_o$ and $b_o$ are its weights and biases, respectively. Here, $\sigma$ is the sigmoid activation function. 

We also compare the LSTM's performance with that of an FNN, namely an MLP (multilayer perceptron). MLPs are a type of fully connected FNN first introduced and popularized by \citet{Rumelhart86}, consisting of an input layer, output layer, and hidden layers in between, where the training process is done through backpropagation. The total input $x_i^{s+1}$ of a neuron $i$ of layer $s+1$ takes the form of 
\begin{equation}\label{eq:input}
    x_i^{s+1} = \sum_{j} h_{ij}^s x_{\sigma j}^s + b_i^{s+1},
\end{equation}
where $h_{ij}^s$ is the weight for neuron $j$ of the previous layer $s$ to neuron $i$ of layer $s+1$ and $b_i^{s+1}$ is the threshold of layer $s+1$. $x_{\sigma j}^s = \sigma (x_i^{s})$ is the output from neuron $j$ from the previous layer $s$, where a nonlinear activation function $\sigma(\cdot)$ is applied to the  input. Most common activation functions include sigmoid, tanh, or ReLU (rectified linear unit), where the ReLU often learns faster in deeper networks.

\subsection{\textit{Prediction Models}}

The first prediction model, Algorithm~\ref{alg:Algo2} (which we will refer to as SD-LSTM), is a prediction procedure that solely uses a nonlinear model (a neural network) to fit the data. The idea is to first train an LSTM for each of the $K$ communities, and then given a new county, we select the corresponding fitted model for prediction from our repertoire with respect to its nearest neighbor county (in demographic variables, not geographical distance).  That is, we apply the nearest neighborhood in demographic variables to classify the new county's community, and use the model for that community to forecast the county's cases. Specifically, for each community $k \in \{1,...,K\}$, we train an LSTM with the data $\{(\boldsymbol{s}_t^i, y_{i,t+l})_{t=1}^{T_i}, \forall i \in \hat G_k\}$ and this depends on the numbers of steps forward, $l$, we are trying to forecast. For simplicity of notation, for community $k$, we denote all such data items for all counties $i \in  \hat G_k$ by $\{(\boldsymbol{s}_t, y_{t+l}), t \in \hat{\cG}_k^l\}$ and the fitted function by $\hat{f}_k^l(\cdot)$.  Now the second part, the prediction, is that given a new county $i'$'s demographic data $\boldsymbol{d}_{i'}$ and social distancing information $\boldsymbol{S}_{i'} = [\boldsymbol{s}_1^{i'}, \boldsymbol{s}_2^{i'},...,\boldsymbol{s}_{T_i'}^{i'}] \in \mathbb{R}^{T_{i'} \times 3}$, we first find its nearest neighbor county $j = \argmin_{j} \|\boldsymbol{d}_{i'} - \boldsymbol{d}_j\|^2$ and its associated community $\hat{Z}_j$ and use its associated prediction model to predict $\hat{y}_{i',t+l} = \hat f_{k'}^l(\boldsymbol{s}_t^{i'}), t = 1,...,T_{i'}$ with $k'=\hat{Z}_j$.  Algorithm~\ref{alg:Algo2} summarizes this method of prediction.

To predict a future event, the above procedure gives a number of prediction methods. For example, to predict tomorrow's outcome, we can use today's social distancing data with $l = 1$, or yesterday's social distancing data with $l=2$, or the day before yesterday's social distancing data with $l = 3$, and so on. As verified later in Figure~\ref{fig:lagandfeature}, it turns out that $l = 4$ is the best choice of lead, which align with the incubation period of the disease.

\begin{algorithm}
  \caption{SD-LSTM: LSTM Prediction}\label{alg:Algo2}
  \textbf{Part I: Training}\\
  \hspace*{\algorithmicindent} \textbf{Input}: The lead $l$
  \begin{algorithmic}[1]
    \For{$k \in \{1,...,K\}$}
        \State Train LSTM $\hat f_k^l(\cdot)$ using the data $\{(\boldsymbol{s}_t,y_{t+l}), t \in \hat{\cG}_k^l\}$.
    \EndFor \\
    \Return fitted LSTMs $\hat f_k^l(\cdot), k = 1,...,K$.
  \end{algorithmic}
  \textbf{Part II: Prediction}\\
  \hspace*{\algorithmicindent} \textbf{Input:} A new county $i'$, $\boldsymbol{d}_{i'} \in \mathbb{R}^q$, $\boldsymbol{S}_{i'} = [\boldsymbol{s}_1^{i'}, \boldsymbol{s}_2^{i'},...,\boldsymbol{s}_{T_i'}^{i'}] \in \mathbb{R}^{T_{i'} \times 3}$, $\boldsymbol{\hat{Z}}$ and $\hat f_k^l (\cdot), k = 1,...,K$ \hspace*{\algorithmicindent} from Part I.
  \begin{algorithmic}[1]
        \State Find county $i'$'s nearest neighbor $j$ = $\argmin_{j} \|\boldsymbol{d}_{i'} - \boldsymbol{d}_j\|^2$.
        \State Select $\hat f_{k'}^l(\cdot),$ where $k' = \hat{Z}_{j}$.
        \For{$t \in \{1,...,T_{i'}\}$}
            \State $\hat{y}_{i',t+l} = \hat f_{k'}^l(\boldsymbol{s}_t^{i'})$.
        \EndFor \\
  \Return $\boldsymbol{\hat{y}}_{i'} = [\hat{y}_{i',1+l},\hat{y}_{i',2+l},...,\hat{y}_{i',T_i+l}]^T \in \mathbb{R}^{T_{i'}}$.
  \end{algorithmic}\label{alg:Algo2}
\end{algorithm}

Algorithm~\ref{alg:Algo3} takes SD-LSTM a step further to include a linear component, namely, fitting the linear model for each county first with residuals from each community then further modeled by an LSTM.  This idea is related to boosting or nonparametric estimation using a parametric start \citet{fan2009local}, resulting in a semi-parametric fit. Again, the objective of the training part is to obtain $K$ fitted models, one for each community, using semi-parametric regression techniques. More specifically, for  county $i$ with lead $l$, we first fit the following linear regression models
\begin{equation}\label{eq:alg3}
    y_{i,t+l} = \alpha_i^l + (\boldsymbol{s}_t^i)^T \boldsymbol{\beta}_i^l + \varepsilon_{i,t+l}, \quad t = 1,...,T_i.
\end{equation}
After fitting the linear regression models for every county $i \in \hat{\cG}_k^l$, we obtain the residuals $\{\hat \epsilon_{i,t+l}, t \in \hat{\cG}_k^l\}$ and save all the coefficients $\alpha_i^l, \boldsymbol{\beta}_{i}^l$ for $i \in \hat{\cG}_k^l, k = 1,...,K$.  
We then extract the information further from $\{(\boldsymbol{s}_t, \hat{\epsilon}_{t+l}), t \in \hat{\cG}_k^l\}$ by fitting an LSTM to obtain the fitted $\hat g_k^l(\cdot)$. Then, for the prediction of the new county $i'$, we follow the same steps as those in SD-LSTM but the final prediction is instead adding the linear fit of the nearest neighbor county and the LSTM fit of the community corresponding to the nearest neighbor county: 
$$
\hat{y}_{i',t+l} = \alpha_{j}^l + (\boldsymbol{s}_t^{i'})^T \boldsymbol{\beta}_{j}^l+ \hat g_{k'}^l(\boldsymbol{s}_t^{i'}),
$$ 
where $k' = \hat Z_j$. We will refer to this model as SD-SP.  The idea is summarized in Algorithm~\ref{alg:Algo3}.

\begin{algorithm}
  \caption{SD-SP: Semi-parametric Prediction}\label{alg:Algo3}
  \textbf{Part I: Training}\\
  \hspace*{\algorithmicindent} \textbf{Input}: The lead $l$
  \begin{algorithmic}[1]
    \For{$k \in \{1,...,K\}$}
         \State Fit the regression models \eqref{eq:alg3} for $i \in \hat{\cG}_k^l$ and obtain the residuals $\{\hat \epsilon_{t+l}, t \in \hat{\cG}_k^l\}$.
        \State Train LSTM using $\{(\boldsymbol{s}_t, \hat{\epsilon}_{t+l}), t \in \hat{\cG}_k^l\}$ 
    \EndFor \\
    \Return fitted LSTMs $\hat g_k^l(\cdot)$ and all $\alpha_{i}^l$ and $\boldsymbol{\beta}_{i}^l$ for $i \in \hat{\cG}_k^l,  k = 1,...,K$. 
  \end{algorithmic}
  \textbf{Part II: Prediction}\\
  \hspace*{\algorithmicindent} \textbf{Input} A new county $i'$, $\boldsymbol{d}_{i'} \in \mathbb{R}^q$, $\boldsymbol{S}_{i'} = [\boldsymbol{s}_1^{i'}, \boldsymbol{s}_2^{i'},...,\boldsymbol{s}_{T_i'}^{i'}] \in \mathbb{R}^{T_{i'} \times 3}$, $\boldsymbol{\hat{Z}}$ and $\hat g_k^l (\cdot)$, $\alpha_{i}^l$ and $\boldsymbol{\beta}_{i}^l$  \hspace*{\algorithmicindent} for $i \in \hat{\cG}_k^l$, $k = 1,...,K$ from Part I.
  \begin{algorithmic}[1]
        \State Find county $i'$'s nearest neighbor $j$ = $\argmin_{j} \|\boldsymbol{d}_{i'} - \boldsymbol{d}_j\|^2$.
        \State Select  $\alpha_{j}^l$ and $\boldsymbol{\beta}_{j}^l$ for county $j$.
        \State Select $\hat g_{k'}^l(\cdot),$ where $k' = \hat{Z}_{j}$.
        \For{$t \in \{1,...,T_{i'}\}$}
            \State $\hat{y}_{i',t+l} = \alpha_{j}^l + (\boldsymbol{s}_t^{i'})^T \boldsymbol{\beta}_{j}^l+ \hat g_{k'}^l(\boldsymbol{s}_t^{i'})$.
        \EndFor \\
    \Return $\boldsymbol{\hat{y}}_{i'} = [\hat{y}_{i',1+l},\hat{y}_{i',2+l},...,\hat{y}_{i',T_i+l}]^T \in \mathbb{R}^{T_{i'}}$.
    \end{algorithmic}
\end{algorithm}

We also include three other algorithms for comparison purposes. The first replaces the LSTM fit $\hat f_k^l(\cdot)$ of community $k$ in SD-LSTM with a linear model. This corresponds to fitting  \eqref{eq:alg3} without further boosting by an LSTM. For simplicity, we shall refer to this approach as the SD-LM (social distancing linear model). The second one is to use both demographic and social distancing data to fit an LSTM. This approach is identical to SD-LSTM, but includes the $q = 7$ significant demographic variables in Table~\ref{Tab:pvaluesR} in addition to the three social distancing variables.  Similarly, we shall refer to this approach as the DSD-LSTM (demographic and social distancing LSTM). DSD-LSTM is expected to improve the performance of Algorithm~\ref{alg:Algo2} due to the additional information from the demographic variables. The final model is similar to SD-LSTM but instead of an LSTM, we use an MLP with two hidden layers (we will refer to this model as SD-MLP).

\subsection{\textit{Implementation}}
For the LSTM, the optimization algorithm used is Adam with a learning rate of 0.01. We also test the performance of various lags to see which yields the highest out-of-sample $R^2$, defined as follows for a given new county $i'$ and lead $l$:
\begin{equation} \label{eq:R2}
    1 - \frac{\sum_{t=1}^{T_{i'}} (y_{i',t+l}-\hat{y}_{i',t+l})^2}{\sum_{t=1}^{T_{i'}} (y_{i',t+l}-\bar{y}_{i', t+l})^2},
\end{equation}
where $y_{i',t+l}$ is the observed value, $\hat{y}_{i',t+l}$ is the predicted value, and $\bar{y}_{i',t+l} = 1/T_{i'} \sum_{t=1}^{T_{i'}} y_{i',t+l}$, serving as the baseline predictor.
The average, median, and standard deviation of the $R^2$ values are then taken across all counties in the testing sample. Additionally, for any model involving an LSTM, up to the minimum length $\Tilde{T} = \min\limits_{i=1,...,N} T_i$ is taken for each county since the LSTM needs each sample to have uniform time steps. Therefore, $T_i = \Tilde{T}$ for each county $i$ in the case of SD-LSTM, SD-SP and the DSD-LSTM model. For information regarding the hidden layers used and input shapes in the neural network models, see Table \ref{Tab:modeldes}.

\begin{table*}[h!]
\centering
\footnotesize
\begin{tabular}{@{}lllclllclll@{}}\toprule 
	& No. of hidden layers & Type and no of nodes & Input shape\\ \midrule
SD-LSTM & 1 &LSTM, 10& $N \times \:\Tilde{T} \times 3$\\
SD-SP & 1 &LSTM, 10& $N \times \:\Tilde{T} \times 3$\\
DSD-LSTM & 1 &LSTM, 10& $N \times \:\Tilde{T} \times 10$\\
SD-MLP & 2 & Dense, 10, 10& $\sum_{i=1}^{N} T_i \times 3$\\
\bottomrule
\end{tabular}
\caption{\footnotesize Number of hidden layers, the type and number of nodes of each hidden layer, and input shape of each model that contains an NN.}
\label{Tab:modeldes}
\vspace{-0.5cm}
\end{table*}

Due to the nature of neural networks and considering the relatively small sample size, we conduct five-fold cross-validation to evaluate the learning models. We divide all the counties into $5$ train-test splits, where the correlation matrix is re-calculated on only the training set. Then, for each $K = 1,...,5$, Algorithm$~\ref{alg:Algo1}$ is executed on the training set for that particular split. Hence, we have $25$ sets of results for each model (five for each of the five train-test splits). 

\subsection{\textit{Data}} \label{preddata}
Social distancing data is courtesy of Unacast and its COVID-19 Social Distancing Scoreboard. The scoreboard tracks mobile device movement and has three metrics that quantify the level of social distancing people in a particular county are practicing. The first metric is the percentage change in total distance traveled, averaged across all devices, compared to a pre-Corona baseline. The second is the percentage change in the number of visitations to non-essential places compared to a pre-Corona baseline. For these two metrics, the pre-Corona baseline of a county on a particular day is defined as the average of the four corresponding pre-weekdays (at least four weeks before the day). For example, for Monday 3/30, the pre-Corona baseline of the first metric is the average of the first metric for the four Mondays: 2/10, 2/17, 2/24, and 3/2. The final metric is the rate of human encounters as a fraction of the pre-Corona national baseline. The pre-Corona national baseline for this metric is the average of the metric taken over four weeks that immediately precede the COVID-19 outbreak (02/10/2020 - 03/08/2020) as defined by Unacast. Since this data starts at 02/25/2020 which is after the start of the Coronavirus cases data (01/22/2020), we perform prediction on the period 02/25/2020 - 7/10/2020, which is the start of the \say{initial phase} until the end of the \say{recent phase}. Also note that not all counties from Johns Hopkins CCSE data and Data Planet are available at Unacast's database so out of the $633$ counties from \autoref{SFdata}, this section is performed on $627$ counties.

\begin{table*}[h!]
\centering
\begin{tabular}{@{}lllclllclll@{}}\toprule 
& \multicolumn{3}{c}{In-Sample} & \multicolumn{3}{c}{Out-of-Sample}\\
\cmidrule{2-3} \cmidrule{4-5} \cmidrule{6-7}
	$Model$& Mean & Median & Std Dev & Mean & Median & Std Dev\\ \midrule
Model 1, $K = 1$ & 0.3447&	0.6376&0.8642&	0.3281&0.6192&0.7978\\
Model 1, $K = 2$ &	0.4733&0.6372&0.5632&	0.2170&	0.5437&1.0514\\
Model 1, $K = 3$&	0.5003&	0.6527&0.5186&	0.2792&	0.5600&	0.8986\\
Model 1, $K = 4$&	0.5211&0.6580&0.4738&0.2485&0.5245&	0.9007\\
Model 1, $K = 5$&0.5113&0.6593&	0.5094&0.2024&0.5461&1.0277\\
Model 2, $K = 1$&-2.0229&-1.8148&0.9800&-2.9663&-1.9824&7.8948\\
Model 2, $K = 2$&-1.8886&-1.7337&0.8901&-2.9613&-1.8356&8.1548\\
Model 2, $K = 3$&-1.8819&-1.6990&0.8471&-2.9393&-1.7641&8.0641\\
Model 2, $K = 4$&-1.8647&-1.7513&	0.8371&-2.9220&-1.7512&8.1715\\
Model 2, $K = 5$&-1.8673&-1.7500&	0.8016&-2.9281&-1.7872&8.0991\\
Model 3, $K = 1$&-0.1227&-0.0271&0.3236&-0.1458&-0.0311&0.4000\\
Model 3, $K = 2$&-0.1148&-0.0260&0.2941&-0.1586&-0.0296&0.5145\\
Model 3, $K = 3$&-0.1138&-0.0252&	0.2936&-0.1563&-0.0276&0.5054\\
Model 3, $K = 4$&-0.1095&-0.0239&0.2771&-0.1659&-0.0302&0.6044\\
Model 3, $K = 5$&-0.0991&-0.0246&0.2559&-0.1542&-0.0317&0.5320\\
Model 4, $K = 1$&0.4855&0.6429&0.5114&0.4513&0.6251&0.5384\\
Model 4, $K = 2$&0.5486&0.6645&0.3786&0.3567&0.5556&0.7031\\
Model 4, $K = 3$&0.5473&0.6554&	0.3952&0.3611&0.5672&0.6536\\
Model 4, $K = 4$&0.5489&0.6522&	0.3930&0.2947&0.4912&0.7717\\
Model 4, $K = 5$&0.5467&0.6550&	0.3812&0.2516&0.5223&0.8706\\
Model 5, $K = 1$&-0.1894&-0.0394&0.4866&-0.2098&-0.0430&0.4551\\
Model 5, $K = 2$&-0.1507&-0.0365&0.3679&-0.1886&-0.0411&0.4241\\
Model 5, $K = 3$&-0.1432&-0.0383&0.3538&-0.1729&-0.0415&0.3932\\
Model 5, $K = 4$&-0.1318&-0.0415&0.3072&-0.1362&-0.0413&0.3277\\
Model 5, $K = 5$&-0.1206&-0.0374&0.2633&-0.1560&-0.0462&0.3741\\
\midrule
\end{tabular}
\caption{02/25/2020 - 7/10/2020 in-sample and out-of-sample $R^2$ for Model 1 (SD-LSTM), Model 2 (SD-SP), Model 3 (SD-LM), Model 4 (DSD-LSTM), and Model 5 (SD-MLP) for $K = 1,2,3,4,5$. The average values for mean, median and standard deviation are taken for each of the $5$ folds. For $K=1$, we assume that all counties belong to one group so we take all counties in the training data to train the neural network. The results are based on $l = 4$ and a five-fold cross-validation. $501$ of the total $627$ counties are used as training data (in-sample) and $126$ counties are used as testing data (out-of-sample).}
\label{Tab:Rsq}
\end{table*}

\begin{figure}[h!]
  \centering
  \includegraphics[width=165mm,height=100mm]{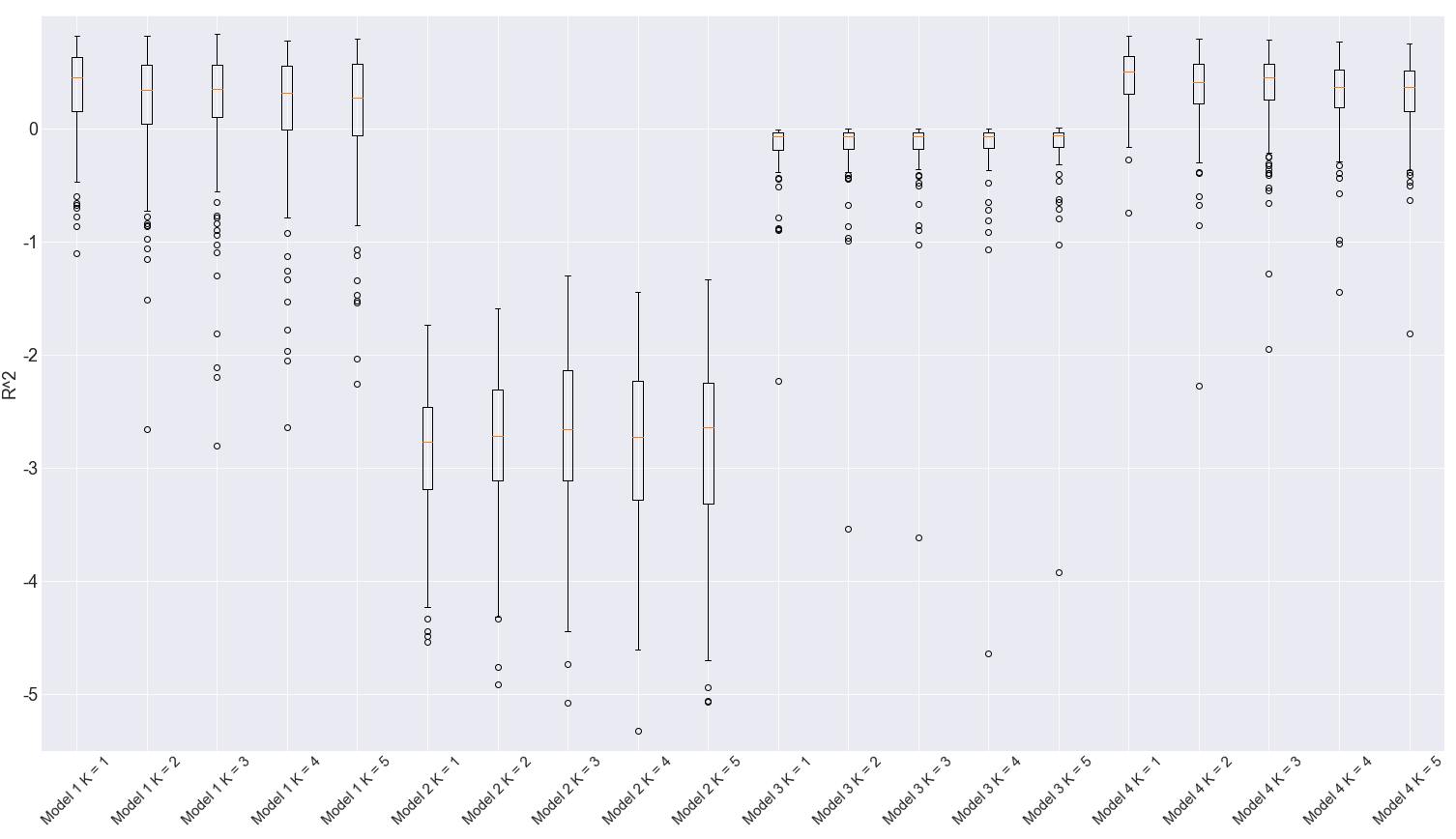}
  \caption{\footnotesize Out-of-sample $R^2$ boxplots for all counties using Model 1 (SD-LSTM), Model 2 (SD-SP), Model 3 (SD-LM) and Model 4 (DSD-LSTM) for $K = 1, 2,3,4,5$. The results are based on $l = 4$ and the period 02/25/2020 - 7/10/2020.}
  \label{fig:ROS}
\end{figure}

\begin{table*}[h!]
\footnotesize
\centering
\begin{tabular}{@{}lllclllclll@{}}\toprule 
& \multicolumn{3}{c}{In-Sample} & \multicolumn{3}{c}{Out-of-Sample}\\
\cmidrule{2-3} \cmidrule{4-5} \cmidrule{6-7}
	$Feature$& Mean & Median & Std Dev & Mean & Median & Std Dev\\ \midrule
Trial 1 &0.2349&0.4842&0.9238&0.4523&0.4956& 0.3505\\
Trial 2 &0.2764&0.5569&0.9678&0.3304&0.4995&0.7916\\
Trial 3&0.2921& 0.5628&0.9442&0.1163&0.4980& 1.4862\\
Trial 4&0.2900&0.5594& 0.9677&0.3410&0.5454& 0.7049\\
Trial 5&0.3404&0.5430&0.8943&-0.0455& 0.4367& 1.3895\\
Median &0.2900&0.5569&	0.9442&0.3304&0.4980&0.7916\\
\bottomrule
\end{tabular}
\caption{02/25/2020 - 7/10/2020 random assignment in-sample and out-of-sample $R^2$ for Model 1 (SD-LSTM), $K = 2$. Each trial is completed via randomly assigning each test county of one of the train-test splits to either community 1 or community 2.}
\label{Tab:Rsq-random}
\end{table*}

\subsection{\textit{Results and Discussion}}

Among the four prediction models we implemented using the county-level social distancing measures (see \autoref{preddata}), for $K=1,2,3$, Model 4 (DSD-LSTM) slightly outperforms Model 1 due to the use of the seven additional demographic variables. Model 1 (SD-LSTM) proves to result in the highest average and median out-of-sample $R^2$ for $K = 4,5$. Models 2 (SD-SP) and 3 (SD-LM) have much poorer performance across the board, which implies that these two models are worse than a horizontal line fit. It is also worth mentioning that the neural network correction part of Model 2 is incredibly hard to tune to be able to outperform the linear model Model 3 on its own. In this case, not only was it not able to enhance Model 3's results, Model 2's correction actually worsened the model's predictive ability. Other nonparametric methods other than a neural network were also used (such as support vector regression) but all had a similar lackluster effect, implying that boosting or enhancing the linear estimator with a nonlinear estimator is not beneficial in this case. Model 1's and Model 4's superiority suggests a nonlinear effect that the LSTM was able to extract, but the linear, semi-parametric, and MLP were unable to do so. 

For Models 1 and 4, stratifying the communities through our method does make a difference in-sample since increasing $K$ improves the models' mean and median in-sample $R^2$. However, this is not the case for out-of-sample as $K = 1$ produces the best results (no heterogeneity) and the out-of-sample $R^2$ continues to drop from $K = 2$ to $5$. It is reasonable to conclude that the decrease in sample size for each community training (e.g. $K = 1$ uses all $501$ counties to train while $K = 5$ uses on average $1/5$th of that number to train each community) is hurting the model's ability to take advantage of the heterogeneity embedded in the communities. Thus, since neural networks have an advantage in large sample size settings, the effect of the reduction in sample size for larger $K$s outweighs the community difference captured by community detection (Algorithm~\ref{alg:Algo1}). We also include Model 5 (an FNN with two hidden layers, each with 50\% dropout) to contrast the LSTM with. The performance is similar to Model 3 in that it is no better than a constant fit. The advantage of the LSTM is highlighted here since the output is dependent on previous computations, unlike the FNN that assumes the inputs (as well as outputs) are independent of each other. As COVID-19 cases are sequential information, the LSTM is clearly preferable to predict with.
See Table \ref{Tab:Rsq} for the detailed breakdown by model and by the number of clusters $K$. Figure \ref{fig:ROS} contains the out-of-sample $R^2$ box plots for the four models with  $K = 1,2,3,4,5$. 

\begin{figure}[h!]
  \centering
  \includegraphics[width=165mm,height=50mm]{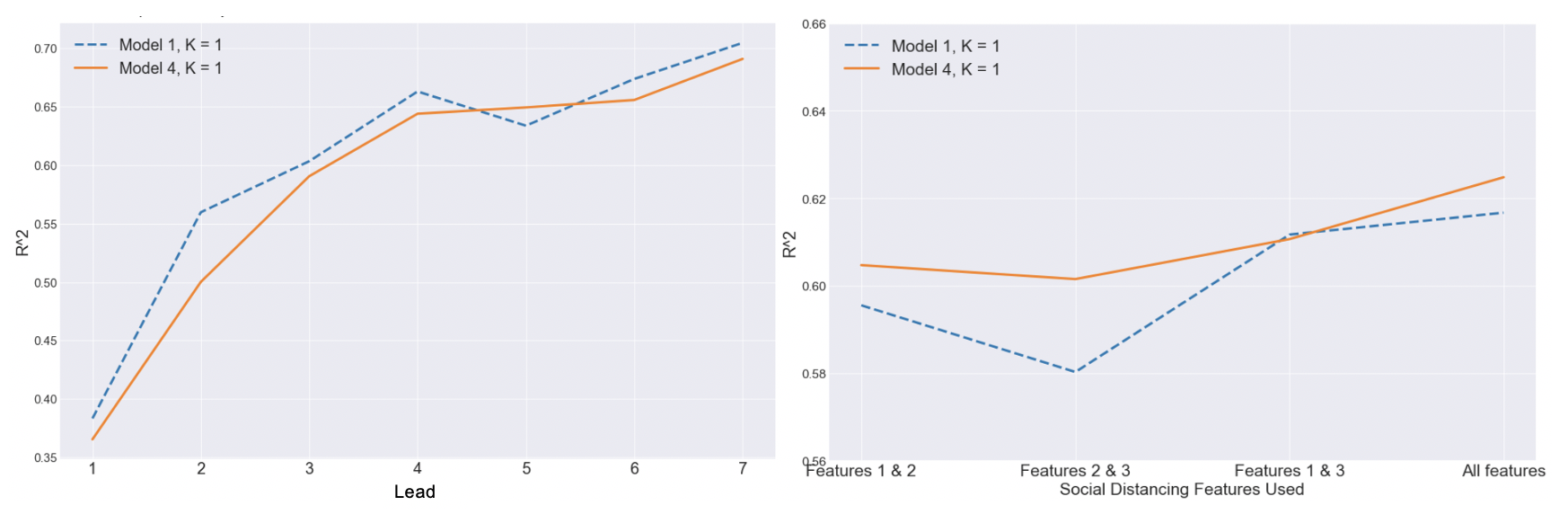}
  \vspace{-0.3in}
  \caption{\footnotesize The left panel is the average out-of-sample $R^2$ for Model 1, $K = 1$ and Model 4, $K = 1$ for $l = 1,2,3,4,5,6,7$ for one train-test split. The right panel is average out-of-sample $R^2$ for the same models, where one social distancing feature is left out each time. Both panels are of the phase 02/25/2020 - 7/10/2020 and based on five-fold cross-validation.}
  \label{fig:lagandfeature}
  \vspace{-0.5cm}
\end{figure}

To ascertain whether using information from community detection still plays a role despite $K = 1$
being the best setting for out-of-sample prediction, we randomly assign each testing county to an existing community instead of using the nearest neighbor method. As shown in Table \ref{Tab:Rsq-random}, after repeating this five times for Model 1, $K = 2$, the median in-sample $R^2$ values are much lower compared to that of the same model in Table \ref{Tab:Rsq} (median of 0.5569 vs 0.6372, respectively). Albeit a smaller difference, the out-of-sample median 0.4980 is also smaller than the 0.5437 in Table \ref{Tab:Rsq}. This demonstrates that community detection can still categorize the nature of different counties' growth trajectories but this effect is likely outweighed by the diminishing sample size as $K$ increases.

Also note that before obtaining the prediction results for each algorithm, the hyperparameter of the appropriate lead was chosen by comparing the average $R^2$ values for each lead. The left panel of Figure \ref{fig:lagandfeature} presents the median out-of-sample $R^2$ vs $l = 1,...,7$ for the two best models Model 1, $K = 1$, and Model 4, $K = 1$ as examples. Since out-of-sample $R^2$ is plateaus after a four-day lead, we fixed $l = 4$ as a larger lead would decrease precision and it is important to be consistent with studies that show the median incubation period of COVID-19 is 4-5 days \citep{Guan20,Lauer20}. Furthermore, anything longer than a week or so is rarely used in epidemiological and sociological studies. In addition, we quantify the social distancing feature importance by averaging the out-of-sample $R^2$ when we leave each feature out one at a time. Evidently, the right panel of Figure \ref{fig:lagandfeature} suggests that although there is no distinct drop in performance, leaving out feature 1 (percent change in total distance traveled) results in the largest decline in $R^2$ whereas leaving out feature 2 (percent change in the number of visitations to non-essential places) results in the smallest decline.

\begin{center} \section{CONCLUSION} \end{center}
By utilizing spectral clustering to recover communities, we develop a framework to detect COVID-19 communities and discover meaningful interpretations of the clusters. We use the correlation matrix instead of the canonical Laplacian as it offers more meaningful insight and more distinct clusters. The resulting communities are distinct in the nature of their respective growth trajectories and there are several demographic variables that further distinguish these growth communities. Singling out the significant demographic features that have explanatory power of a county's growth community membership, we discover that not all of these variables are intuitive when it comes to their role in impacting COVID-19 cases. 

After modeling and interpreting historical disease progression, we turn to study future growth trajectories by incorporating social distancing information. We are able to reliably predict the logarithmic trends in case growth through the use of LSTMs and also verify that the counties are far from homogeneous - the obtained communities contain crucial information necessary for prediction in-sample. As for the LSTM's out-of-sample predictive power, the effect of the decline in sample size when increasing the stratification of counties into more communities dominates the heterogeneity between counties' growth curves that community detection uncovers. However, after comparing results to randomly assigning counties to different communities, the method we propose still demonstrates that using the community detection results boosts the models' predictive performance.
 
We do, therefore, acknowledge that there could be other latent features that we did not capture in this study and that the three social distancing metrics used here may not paint the complete picture. Furthermore, we do not address the effect of government intervention at given time points that may have altered the disease progression. These could all be points that can be further investigated. Despite these potential shortcomings, the analysis conducted on the first phase of the disease here can also be compared to the second phase, which we are currently experiencing. As the U.S. and many other countries are witnessing an even more extraordinary uptick in cases again, we foresee several possible future applications of our study, including to other contagious disease outbreaks. Another interesting future work 
is to utilize the confidence distribution framework \citet{xie2011confidence} to combine  studies from independent data sources from different countries. 
\subsection*{Acknowledgements}
We would like to thank Unacast Inc. for providing us with their extensive social distancing data. 
The work was in part supported by NSF Grants DMS-2013789,  DMS-1712591 and DMS-2034022, and NIH grant 2R01-GM072611-15.

\singlespacing
\bibliographystyle{plainnat}
\bibliography{references}

\newpage
\setcounter{page}{1}
\section*{Supplementary Material: Clustering with adjacency matrices}\label{appendixa}

Algorithm~\ref{alg:Algo4} outlines the spectral clustering procedure with adjacency matrices $\boldsymbol{A}_1$ and $\boldsymbol{A}_2$ and Tables \ref{Tab:A1} and \ref{Tab:A2} present Groups 1 and 2's mean, median, and standard deviation of the 17 features.

\begin{algorithm}
  \caption{Normalized Laplacian spectral clustering}\label{alg:Algo4}
  \textbf{Input} Similarity matrix $\boldsymbol{S} \in \mathbb{R}^{n \times n}$ and $K$ clusters to obtain.
  \begin{algorithmic}[1]
        \State Obtain the adjacency matrix $\boldsymbol{A}$.
        \State Compute the normalized, symmetric Laplacian matrix $\boldsymbol{L} = \bI -\boldsymbol{D}^{-1/2}\boldsymbol{AD}^{-1/2}$.
        \State Compute the smallest $K$ eigenvectors in absolute value $\boldsymbol{u}_1,..., \boldsymbol{u}_K$ of $\boldsymbol{L}$ and construct $\boldsymbol{\hat{U}} \in \mathbb{R}^{n \times K}$ be the matrix with the eigenvectors as columns.
        \State Normalize rows of $\boldsymbol{\hat{U}}$ to have unit norm 1 to get $\boldsymbol{\hat{U}}_{norm}$. 
        \State Clusters the rows of $\boldsymbol{\hat{U}}_{norm}$ with $k$-means.
  \end{algorithmic}
  \textbf{return} Partition $\hat{G}_1, . . . , \hat{G}_K$ of the nodes. \label{alg:Algo4}
\end{algorithm}

\begin{table*}[h!]
\centering
\footnotesize
\begin{tabular}{@{}lllclllclll@{}}\toprule 
& \multicolumn{3}{c}{Group 1} & \multicolumn{3}{c}{Group 2}\\
\cmidrule{2-3} \cmidrule{4-5} \cmidrule{6-7}
	$Feature$& Mean & Median & Std Dev & Mean & Median & Std Dev\\ \midrule
Population Density & 501.873 & 191.700 & 1038.69 & 693.482&210.820 &3575.14\\
Median Household Income & 56442.4& 53812.0 &14879.8& 56872.8& 54539.0& 15776.4\\
$\%$ Poverty & 13.8581& 13.3000 &5.63048&13.6742& 12.6500& 5.92796\\
$\%$ 1-person households &26.9710& 27.2453& 4.37449& 27.1832& 27.5257& 4.39231\\
$\%$ 5 or more person households &9.25104& 8.59139& 3.24107& 9.05622& 8.56888& 3.02516\\
$\%$ households w 60 y/o and older &38.7766& 38.8296& 6.56590&39.2932&39.0888&6.33139\\
$\%$ w low access to stores &20.6451& 20.5700& 9.36471& 22.0766& 22.1900& 10.0694\\
$\%$ low income w low access to stores &6.77756& 6.14000&4.23293& 7.29463&6.42000& 4.78238\\
$\%$ households w low access to stores &2.42279&2.03000& 1.53641&2.50819&2.22000&1.63217\\
25 y/o and older w Bachelor's /1,000 &116.308&112.763& 43.5637& 117.340& 112.172& 41.5866\\
$\%$ White &78.6730& 82.7486& 15.2783&77.9016&81.8775& 16.9468\\
$\%$ Black &11.8118& 6.07323& 14.0567& 13.0591& 6.64656& 15.9400\\
$\%$ Asian &2.78995& 1.41062& 3.98255& 2.76574&1.49532& 4.19016\\
No of bars&47.7015& 20.0000& 90.2649&38.6136& 18.0000& 60.4562\\
No of grocery stores&89.3181&30.0000& 211.193& 73.2670& 29.0000& 130.915\\
No of restaurants&14.9178& 9.00000& 15.9277& 13.2853& 8.00000& 14.1468\\
$\%$ take public transportation&0.70840& 0.24056& 1.68443& 0.96613& 0.24533& 6.04509\\
\bottomrule
\end{tabular}
\caption{\small $\boldsymbol{A}_1$ clusters' mean and median values for selected features for each community $K=2$. Model $\boldsymbol{A}_1$ corresponds to Algorithm~\ref{alg:Algo4} where we use the $k$-nearest neighbors graph ($k=7$) as $\boldsymbol{A}$.}
\label{Tab:A1}
\end{table*}

\begin{table*}[h!]
\centering
\footnotesize
\begin{tabular}{@{}lllclllclll@{}}\toprule 
& \multicolumn{3}{c}{Group 1} & \multicolumn{3}{c}{Group 2}\\
\cmidrule{2-3} \cmidrule{4-5} \cmidrule{6-7}
	$Feature$& Mean & Median & Std Dev & Mean & Median & Std Dev\\ \midrule

Population Density & 712.520&211.600& 3570.04& 483.554&189.310& 1076.73\\
Median Household Income & 57683.7& 54873.0 &15640.9& 55628.1& 53739.0 &14959.5\\
$\%$ Poverty &13.3879&12.2500& 5.92222& 14.1455&13.4000& 5.61381\\
$\%$ 1-person households &27.0335&27.2366& 4.36119& 27.1226& 27.5300& 4.40790\\
$\%$ 5 or more person households &9.06375& 8.58968& 2.95150& 9.24261& 8.57241& 3.30828\\
$\%$ of households w 60 y/o and older &38.7801& 38.3954& 6.74182&39.2953& 39.3441& 6.13981\\
$\%$ w low access to stores &21.7977& 21.9050& 9.85552& 20.9322& 20.5100& 9.62812\\
$\%$ low income w low access to stores &7.09803& 6.40000& 4.67050& 6.97775& 6.13000& 4.37390\\
$\%$ households w low access to stores &2.40026& 2.20000& 1.51320& 2.53181& 2.11000& 1.65335\\
25 y/o and older w Bachelor's /1,000 &119.289& 113.846& 44.2816& 114.350& 109.250& 40.6497\\
$\%$ White &78.9164& 82.9167& 15.8005&77.6482& 81.4956& 16.4603\\
$\%$ Black &12.1216& 6.12919& 14.6064& 12.7612& 6.38394& 15.4733\\
$\%$ Asian &2.73678& 1.51268& 3.88658& 2.81899& 1.41062& 4.28174\\
No of bars&41.2933& 17.0000& 79.4118& 44.9514&21.0000& 73.3541\\
No of grocery stores&79.5744& 29.0000& 170.279& 82.6123& 30.0000& 179.320\\
No of restaurants&14.2222& 8.00000& 15.6239& 13.9210& 8.00000& 14.4344\\
$\%$ take public transportation&0.95377& 0.23521& 5.91278&0.71807&0.24792& 2.01788\\
\bottomrule
\end{tabular}
\caption{\footnotesize $\boldsymbol{A}_2$ clusters' mean and median values for selected features for each community $K=2$. Model $\boldsymbol{A}_2$ corresponds to Algorithm~\ref{alg:Algo4} where we use the $\epsilon$-neighborhood graph ($\epsilon = 0.007$). Group 1 and 2 are the obtained partitions $\hat{G}_1$ and $\hat{G}_2$, respectively.}
\label{Tab:A2}
\end{table*}

\end{document}